\documentclass[11pt,a4paper,onecolumn]{article}

\usepackage[T1]{fontenc}
\usepackage{times}
\usepackage[margin=25mm]{geometry}
\usepackage{color,array,amsthm,amsmath,amssymb}
\usepackage{graphicx}
\usepackage[hidelinks]{hyperref}

\theoremstyle{definition}
\newtheorem{problem}{Problem}
\newenvironment{keywords}{\par\noindent\small\textbf{Keywords:}\space\ignorespaces}{\par}
\newenvironment{authorbiography}[2][]{
  \par\addvspace{12pt}\noindent
  \begin{minipage}{\linewidth}
  \begin{minipage}[t]{0.17\linewidth}
  \vspace{0pt}#1
  \end{minipage}\hfill
  \begin{minipage}[t]{0.79\linewidth}
  \vspace{0pt}\small\textbf{#2}\space\ignorespaces
}{
  \end{minipage}
  \end{minipage}\par
}

\let\originalthebibliography\thebibliography
\renewcommand{\thebibliography}[1]{
  \originalthebibliography{#1}\small
  \setlength{\itemsep}{5pt}\setlength{\parsep}{0pt}
}
\title{Temporal Attention-Enhanced Mixture-of-Experts Guidance for Multi-Revolution Low-Thrust Rendezvous}
\author{\normalsize Chenglong Wu, Yi Zhou, Hongxu Zhu, Jialong Song, and Hexi Baoyin\thanks{Corresponding author: Hexi Baoyin. E-mail: \href{mailto:baoyin@tsinghua.edu.cn}{baoyin@tsinghua.edu.cn}. This work was supported by the National Natural Science Foundation of China under Grant 12602076.}\\[4pt]
\small School of Aerospace Engineering, Tsinghua University\\
\small Beijing 100084, China}
\date{}
\hypersetup{pdftitle={Temporal Attention-Enhanced Mixture-of-Experts Guidance for Multi-Revolution Low-Thrust Rendezvous},pdfauthor={Chenglong Wu, Yi Zhou, Hongxu Zhu, Jialong Song, Hexi Baoyin}}

\begin{document}
\maketitle

\begin{abstract}
  Multi-revolution low-thrust rendezvous requires both long-duration orbital
  transfer and precise terminal control. This paper proposes a temporal
  attention-enhanced mixture-of-experts guidance law trained by deep
  reinforcement learning. A far-range expert acts on modified equinoctial
  elements, a terminal expert acts on the relative state in the target's local
  frame, and a learned gate blends their thrust commands. The terminal expert
  encodes its observation history with a Transformer to address partial
  observability, and the gate uses a separate encoder to account for the
  resulting history-dependent command. The experts are pretrained separately,
  the gate is trained with both experts frozen, and the complete guidance law is
  then fine-tuned end to end. On 600 test missions, it achieves rendezvous in
  every case at tolerances of 10~m and 1~m/s, reduces mean flight time and
  $\Delta v$ by 50.6\% and 44.5\% relative to distance-based hard switching, and
  remains successful without retraining under navigation errors, unmodeled
  orbital perturbations, eclipse-induced thrust outages, and thrust loss.
\end{abstract}


\section{INTRODUCTION}
Autonomous rendezvous is a key enabling capability for on-orbit servicing and
active debris removal \cite{hakima2018deorbiter,bucci2016analytical}. Electric
propulsion offers high propellant efficiency, but its small thrust acceleration
spreads orbital reconfiguration over many revolutions. Unlike orbit transfer,
rendezvous must also match the target's orbital phase, driving relative
position and velocity to zero simultaneously. Over such long durations,
perturbations accumulate and the actual trajectory departs from a precomputed
plan, motivating closed-loop guidance that carries the chaser from large
initial orbital differences to precise terminal rendezvous.

Low-thrust transfer and rendezvous are commonly formulated as nonlinear
optimal-control problems \cite{betts1998survey,morante2021survey} and solved by
direct methods \cite{hargraves1987direct,zuiani2012direct}, indirect methods
with homotopy and multiple shooting \cite{jiang2012practical,meng2019lowthrust},
or, for multi-revolution missions, techniques that exploit orbital structure
\cite{e2020multirevolution,wu2023atlas,hu2025rapid,zou2025sparse}. Repeatedly
solving such long-horizon problems online is computationally demanding.
Lyapunov-based laws such as the Q-law avoid online optimization
\cite{petropoulos2005refinements} but are sensitive to parameter tuning and are
designed for orbit transfer rather than rendezvous, while sequential convex
optimization \cite{benedikter2019convex} and computationally efficient model
predictive control \cite{larsen2019computationally} reduce, but do not
remove, the cost of iterative online optimization. This motivates learning a
feedback policy offline, so that onboard guidance reduces to evaluating the
learned policy using current observations and, when needed, their recent history.

Neural networks trained on optimized trajectories have enabled real-time
low-thrust transfer control \cite{izzo2021realtime,cheng2019realtime},
missed-thrust recovery \cite{rubinsztejn2020neural}, and certified
close-proximity guidance \cite{wang2026certified}, but constructing
representative training sets can require extensive trajectory optimization.
Deep reinforcement learning instead learns policies through interaction
with the environment, without optimal-control labels, and has been applied to
robust low-thrust trajectory design
\cite{zavoli2021reinforcement,hu2023densely,gao2026robust}, multi-body
low-thrust guidance \cite{lafarge2021autonomous}, proximity operations with
collision avoidance \cite{federici2021deep,qu2022spacecraft}, and
$J_2$-perturbed far-distance rendezvous with continuous high thrust
\cite{di2025trajectory}.

Multi-revolution low-thrust rendezvous is challenging for a single policy,
which must handle both long-duration transfer across large orbital differences
and precise terminal regulation, two regimes whose state scales and control
priorities differ markedly. Single-policy approaches have been developed for
fixed-time rendezvous between coplanar circular orbits \cite{federici2022meta}
and for asteroid approach using zero-effort-miss/zero-effort-velocity feedback
\cite{yang2024robust}. Decomposing control into multiple policies has
also been explored: Zaidi et al. cascade two soft actor--critic (SAC)~\cite{haarnoja2018soft} policies for
multi-revolution orbit raising \cite{zaidi2023cascaded}, and hierarchical
reinforcement learning has coordinated multiphase docking
\cite{aborizk2024multiphase} and short-range rendezvous with attitude control
\cite{tammam2025hierarchical}. However, the cascade targets orbit transfer,
which lacks phase and terminal relative-velocity requirements, and the
hierarchical methods address proximity operations, where relative motion is
small and naturally described in a local frame.

Such local descriptions are ill-suited when the initial orbital differences are
large: the Cartesian relative state then reaches large magnitudes and varies
periodically within each revolution, masking the slow secular change produced
by low thrust, and linearized relative-motion models are valid only for small
separations \cite{clohessy1960terminal,schaub2018analytical}. The orbital elements
describing orbit size, shape, and plane evolve slowly under low thrust;
together with a separate phase residual, they reveal the mismatch to be
removed \cite{battin1999astrodynamics}. Near
rendezvous the situation reverses: errors at the level of the terminal
tolerances correspond to minute, coupled residuals in the elements, whereas the
local relative state measures them directly. This paper therefore assigns the
two regimes to separate policies with different state representations: a
far-range policy that observes orbital elements and a terminal policy that
observes the local relative state. By contrast, both policies in the cascade of
\cite{zaidi2023cascaded} operate on orbital elements.

The remaining question is how to allocate control authority between the two
policies. Common allocation rules restrict the weights to narrow families:
distance-based hard switching permits only binary weights set by a range
threshold, and fixed-weight blending applies the same weights throughout the
mission. A continuous, state-dependent weight instead allows both policies to contribute over a transition region, supporting gradual handovers.
Mixture-of-experts (MoE) architectures realize such input-dependent weighting
by combining specialized models through a learned gating network
\cite{jacobs1991adaptive} and have been adopted in reinforcement learning
\cite{yang2020multitask,obandoceron2024mixtures,willi2024mixture}. In the
proposed architecture, a gating network (the gate) blends the thrust commands
of the far-range and terminal experts. Because the allocation is
learned rather than prescribed, the gate can better coordinate the two experts
based on the mission state.

The terminal expert's local observation introduces partial observability.
The relative state and the target mean motion do not specify target-orbit
quantities, such as eccentricity and orbital position, that affect the relative
motion under the perturbed dynamics. The terminal expert therefore faces a
partially observable Markov decision process (POMDP)
\cite{kaelbling1998planning}. To address this partial observability, it uses a
finite window of recent observations, whose evolution provides information
absent from a single frame. Its thrust command then depends on this window,
so the gate must account for the terminal expert's memory when predicting the
effect of a chosen blending weight. We therefore also give the gate an
observation window that retains the history on which the terminal command
depends.

Recurrent policies are established
baselines for learning from observation histories \cite{ni2022recurrent}, and
long short-term memory networks have been applied to spacecraft guidance
\cite{federici2022meta,kim2024timeshift}. Transformers apply content-dependent
attention across sequence elements
\cite{vaswani2017attention,parisotto2020stabilizing} and have been used to
warm-start rendezvous trajectory optimization \cite{guffanti2024transformers},
to guide close-range rendezvous \cite{mohseni2026physics}, and to optimize
low-thrust trajectories \cite{zaidi2025attention,haq2026attention}. These
works use sequence information for guidance or trajectory optimization. For multi-revolution rendezvous, the unresolved issue is how to coordinate experts with different state representations when one expert's command depends on an observation history.

Training the experts and gate jointly from scratch couples their learning
problems: changes in the experts alter the effect of each blending weight,
and the initial far-range policy may provide few useful terminal encounters.
Pretraining the components
separately avoids this coupling, in the spirit of composing pretrained
primitive policies \cite{peng2019mcp} and coordinating existing subcontrollers
\cite{aborizk2024multiphase}; subsequent end-to-end fine-tuning then adapts the
components to each other under the mission-level return.

Motivated by these considerations, this paper proposes a temporal attention-enhanced MoE guidance framework for multi-revolution low-thrust rendezvous. The main contributions are as follows:
\begin{enumerate}
\item An MoE guidance architecture is proposed that resolves the conflicting state representations of the two regimes: a far-range expert observes global states expressed in modified equinoctial elements, a terminal expert observes the local relative state, and a learned gate blends their thrust commands. Each expert thus operates in the representation suited to its regime, and the fixed handover rule is replaced by a learned, state-dependent allocation of control authority, substantially reducing flight time and $\Delta v$ relative to distance-based hard switching.
\item Temporal attention is introduced into both the terminal expert and the gate to handle the partial observability of the local relative state: the expert encodes its observation history to recover target-orbit information absent from a single frame, and the gate encodes a matching history window so that its allocation is conditioned on the same history that determines the terminal command. This markedly improves terminal convergence over a memoryless policy.
\item A three-stage training procedure is developed that decomposes the coupled learning problem into tractable subproblems: the experts are pretrained separately, with the terminal expert started from handover states along far-range trajectories; the gate is trained with both experts frozen; and the complete guidance law is fine-tuned end to end. This decomposition avoids the nonstationary blending targets and scarce terminal encounters of training the full architecture from scratch, substantially reducing the training difficulty.
\end{enumerate}

Section~\ref{sec:problem} formulates the perturbed low-thrust rendezvous problem.
Section~\ref{sec:method} presents the MoE guidance architecture.
Section~\ref{sec:training} describes the staged training and joint fine-tuning
procedure. Section~\ref{sec:results} presents numerical experiments, including
comparisons with conventional methods
and paired robustness assessments.
Section~\ref{sec:conclusion} concludes the paper.

\section{MULTI-REVOLUTION LOW-THRUST RENDEZVOUS PROBLEM}
\label{sec:problem}

\subsection{Mission Scenario}
\label{sec:mission}
\begin{figure}[!htbp]
\centering
\includegraphics[width=0.70\linewidth]{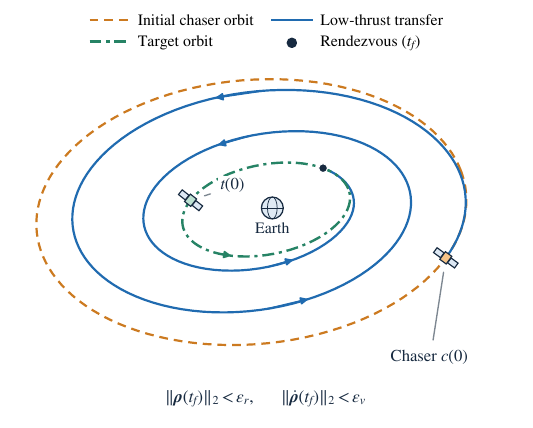}
\caption{Schematic of multi-revolution low-thrust rendezvous between non-coplanar elliptical orbits.}
\label{fig:mission_rendezvous_scenario}
\end{figure}

A low-thrust chaser, indexed by $c$, is required to rendezvous with a passive target, indexed by $t$, in low Earth orbit (Fig.~\ref{fig:mission_rendezvous_scenario}). The two spacecraft start on non-coplanar elliptical orbits with an arbitrary phase difference, and no intermediate orbit, handover point, or phasing strategy is prescribed. Because the thrust acceleration is orders of magnitude smaller than the gravitational acceleration, the transfer spans many revolutions and contains two regimes: a far-range regime of orbital reconfiguration and phasing, in which the separation can reach a sizable fraction of the orbit circumference, and a terminal regime, in which tight position and velocity tolerances require precise control of the relative motion. The two regimes differ by several orders of magnitude in spatial scale and required control resolution. The orbit ranges, thrust level, and guidance interval are given in Section~\ref{sec:results}.

\subsection{Perturbed Dynamics in Modified Equinoctial Elements}
The nominal dynamics used for policy training and baseline evaluation model both spacecraft as point masses under Earth's central gravity and the $J_2$ zonal perturbation. The target is unactuated, and the chaser carries a throttleable low-thrust engine whose commanded direction is realized without attitude lag. $J_2$ is retained because it is the dominant perturbation in low Earth orbit and, over a multi-revolution transfer, induces differential precession of the node and argument of latitude between the two orbits that the guidance law must absorb; it is therefore part of the dynamics from the outset rather than a disturbance. Atmospheric drag, solar radiation pressure, lunisolar third-body gravity, navigation errors, eclipse-induced thrust outages, and reductions in maximum thrust acceleration are omitted from the nominal model. Their effects are considered in the robustness tests of Section~\ref{sec:res_robustness}.

Following the variation-of-parameters formulation of perturbed orbital motion \cite{battin1999astrodynamics}, the osculating orbit of each spacecraft is represented by the modified equinoctial elements (MEEs), which avoid the singularities of classical elements for circular and equatorial orbits \cite{walker1985modified,walker1986erratum}:
\begin{equation}
  \boldsymbol{x}
  =
  \begin{bmatrix}
    p & f & g & h & k & L
  \end{bmatrix}^{\mathrm T},
  \label{eq:mee_state}
\end{equation}
where
\begin{equation}
  \begin{split}
    p&=a(1-e^2),\\
    f&=e\cos(\omega+\Omega),\qquad
    g=e\sin(\omega+\Omega),\\
    h&=\tan(i/2)\cos\Omega,\qquad
    k=\tan(i/2)\sin\Omega,\\
    L&=\Omega+\omega+\nu,
  \end{split}
  \label{eq:mee_definition}
\end{equation}
and $a$, $e$, $i$, $\Omega$, $\omega$, and $\nu$ denote the semimajor axis, eccentricity, inclination, right ascension of the ascending node, argument of perigee, and true anomaly, respectively.

\begin{figure}[!htbp]
\centering
\includegraphics[width=0.62\linewidth]{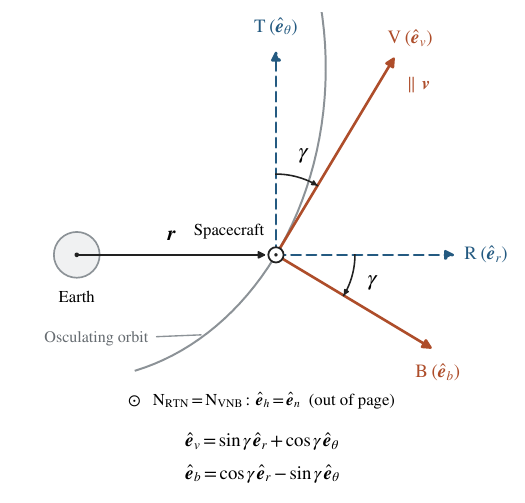}
\caption{RTN (dashed) and VNB (solid) frames with flight-path angle $\gamma$.}
\label{fig:rtn_vnb_geometry}
\end{figure}

Two local orbital frames, centered on a spacecraft and rotating with its orbital motion, are used. The radial--transverse--normal (RTN) frame has basis vectors
\begin{equation}
  \begin{aligned}
    \hat{\boldsymbol{e}}_r&=\frac{\boldsymbol{r}}{\|\boldsymbol{r}\|},\qquad
    \hat{\boldsymbol{e}}_h=\frac{\boldsymbol{r}\times\boldsymbol{v}}{\|\boldsymbol{r}\times\boldsymbol{v}\|},\\
    \hat{\boldsymbol{e}}_\theta&=\hat{\boldsymbol{e}}_h\times\hat{\boldsymbol{e}}_r,
  \end{aligned}
  \label{eq:rtn_basis}
\end{equation}
where $\boldsymbol{r}$ and $\boldsymbol{v}$ are the position and velocity expressed in the Earth-centered inertial (ECI) frame. The velocity--normal--binormal (VNB) frame has basis vectors
\begin{equation}
  \hat{\boldsymbol{e}}_v=\frac{\boldsymbol{v}}{\|\boldsymbol{v}\|},\qquad
  \hat{\boldsymbol{e}}_n=\hat{\boldsymbol{e}}_h,\qquad
  \hat{\boldsymbol{e}}_b=\hat{\boldsymbol{e}}_v\times\hat{\boldsymbol{e}}_n .
  \label{eq:vnb_basis}
\end{equation}
The two frames share the orbit-normal axis. Their in-plane axes are related by the flight-path angle $\gamma$, illustrated in Fig.~\ref{fig:rtn_vnb_geometry} and defined in \eqref{eq:vnb_to_rtn}; for a circular orbit, $\hat{\boldsymbol{e}}_v=\hat{\boldsymbol{e}}_\theta$ and $\hat{\boldsymbol{e}}_b=\hat{\boldsymbol{e}}_r$. The RTN frame is the natural frame for the Gauss variational equations, whereas the VNB frame is used for the thrust command because its first axis is aligned with the direction of maximum orbital-energy rate.

Let $\boldsymbol{a}=[a_r,a_\theta,a_h]^{\mathrm T}$ be the total perturbing acceleration resolved in the RTN frame. The MEE dynamics are \cite{battin1999astrodynamics,walker1985modified,walker1986erratum}
\begin{equation}
  \dot{\boldsymbol{x}}
  =\boldsymbol{B}(\boldsymbol{x})\boldsymbol{a}
  +\boldsymbol{d}(\boldsymbol{x}),
  \label{eq:mee_dynamics}
\end{equation}
where, with $q=1+f\cos L+g\sin L$, $s^2=1+h^2+k^2$, and $\varsigma=h\sin L-k\cos L$,
\begingroup
\setlength{\arraycolsep}{2pt}
\setlength{\medmuskip}{2mu}
\setlength{\thickmuskip}{3mu}
\begin{equation}
  \boldsymbol{B}(\boldsymbol{x})
  =\sqrt{\frac{p}{\mu}}
  \begin{bmatrix}
    0 & \dfrac{2p}{q} & 0\\
    \sin L & \dfrac{(q+1)\cos L+f}{q}
    & -\dfrac{g\varsigma}{q}\\
    -\cos L & \dfrac{(q+1)\sin L+g}{q}
    & \dfrac{f\varsigma}{q}\\
    0&0&\dfrac{s^2\cos L}{2q}\\
    0&0&\dfrac{s^2\sin L}{2q}\\
    0&0&\dfrac{\varsigma}{q}
  \end{bmatrix}
  \label{eq:mee_control_matrix}
\end{equation}
\endgroup
and
\begin{equation}
  \boldsymbol{d}(\boldsymbol{x})
  =
  \begin{bmatrix}
    0&0&0&0&0&\sqrt{\mu p}\left(\dfrac{q}{p}\right)^2
  \end{bmatrix}^{\mathrm T}.
  \label{eq:mee_keplerian}
\end{equation}

The acceleration acting on the chaser is the sum of the thrust acceleration and the $J_2$ perturbation,
\begin{equation}
  \boldsymbol{a}_c
  =a_{\max}\boldsymbol{C}_{\mathrm{VNB}}^{\mathrm{RTN}}
  \boldsymbol{u}+\boldsymbol{a}_{J_2,c},
  \label{eq:chaser_acceleration}
\end{equation}
where $a_{\max}$ is the maximum thrust acceleration and $\boldsymbol{u}$ is the normalized thrust command expressed in the chaser VNB frame, constrained to the admissible set
\begin{equation}
  \mathcal{U}=\{\boldsymbol{u}\in\mathbb{R}^3:\|\boldsymbol{u}\|_2\leq 1\},
  \label{eq:control_constraint}
\end{equation}
and $\boldsymbol{C}_{\mathrm{VNB}}^{\mathrm{RTN}}$ is the direction-cosine matrix from VNB to RTN, whose columns are the VNB basis vectors \eqref{eq:vnb_basis} expressed in RTN components,
\begin{equation}
  \begin{gathered}
    \boldsymbol{C}_{\mathrm{VNB}}^{\mathrm{RTN}}
    =
    \begin{bmatrix}
      \sin\gamma & 0 & \cos\gamma\\
      \cos\gamma & 0 & -\sin\gamma\\
      0 & 1 & 0
    \end{bmatrix},
    \\
    \tan\gamma=\frac{\boldsymbol{r}\cdot\boldsymbol{v}}{\|\boldsymbol{r}\times\boldsymbol{v}\|},
  \end{gathered}
  \label{eq:vnb_to_rtn}
\end{equation}
where $\gamma$ is the flight-path angle of the chaser. The command is piecewise constant: $\boldsymbol{u}(t)=\boldsymbol{u}_k$ for $t\in[t_k,t_{k+1})$, with $t_k=k\Delta t$. Because the command is held in the rotating VNB frame, the inertial thrust direction rotates with the orbit during each interval. The target satisfies \eqref{eq:mee_dynamics} with $\boldsymbol{a}_t=\boldsymbol{a}_{J_2,t}$.

The $J_2$ acceleration is expressed directly in RTN components as a function of the MEEs, which avoids converting to Cartesian coordinates at every integration step. With $r=p/q$ and $\psi=h\cos L+k\sin L$ \cite{betts2010practical},
\begin{equation}
  \boldsymbol{a}_{J_2}
  =-\frac{3J_2\mu R_E^2}{2r^4}
  \begin{bmatrix}
    1-\dfrac{12\varsigma^2}{s^4}\\[1.2ex]
    \dfrac{8\varsigma\psi}{s^4}\\[1.2ex]
    \dfrac{4\left(1-h^2-k^2\right)\varsigma}{s^4}
  \end{bmatrix},
  \label{eq:j2_acceleration}
\end{equation}
where $\mu$, $R_E$, and $J_2$ are the gravitational parameter, equatorial radius, and second zonal harmonic of the Earth. The expression is the RTN projection of the standard inertial $J_2$ acceleration \cite{vallado2013fundamentals}. Equations \eqref{eq:mee_dynamics}--\eqref{eq:j2_acceleration} define the nominal dynamics of both spacecraft used in training.

\subsection{Global and Local State Representations}
\label{sec:frames}
The rendezvous state admits two complementary descriptions.

\emph{Global description.} The orbital geometry and phase mismatch between the two spacecraft are described by the MEE residuals $\Delta\alpha=\alpha_t-\alpha_c$ for $\alpha\in\{p,f,g,h,k\}$ and by the wrapped true-longitude residual
\begin{equation}
  \Delta L
  =\big((L_t-L_c)+\pi\big)\bmod 2\pi-\pi,
  \label{eq:longitude_residual}
\end{equation}
which takes values in $[-\pi,\pi)$ and identifies longitudes differing by $2\pi$.

\emph{Local description.} Terminal accuracy is defined by the relative position and velocity in the target-centered local-vertical/local-horizontal (LVLH) frame, which coincides with the RTN frame \eqref{eq:rtn_basis} of the target: the $x$ axis points radially outward along $\hat{\boldsymbol{e}}_{r,t}$, the $z$ axis is along the target orbit normal $\hat{\boldsymbol{e}}_{h,t}$, and the $y$ axis $\hat{\boldsymbol{e}}_{\theta,t}$ completes the right-handed triad in the along-track direction. The direction-cosine matrix from ECI to LVLH, $\boldsymbol{C}_{I}^{L}$, has these three unit vectors as its rows.
The relative-velocity transformation uses
$\boldsymbol{\omega}^{(0)}_{L/I}=\|\boldsymbol{r}_t\times\boldsymbol{v}_t\|\,\hat{\boldsymbol{e}}_{h,t}/\|\boldsymbol{r}_t\|^2$,
neglecting the contribution from rotation of the target orbit plane \cite{schaub2018analytical}. The local quantities are
\begin{equation}
  \begin{split}
    \boldsymbol{\rho}
    &=\boldsymbol{C}_{I}^{L}(\boldsymbol{r}_c-\boldsymbol{r}_t),\\
    \dot{\boldsymbol{\rho}}
    &=\boldsymbol{C}_{I}^{L}\bigl[(\boldsymbol{v}_c-\boldsymbol{v}_t)\\
    &\qquad{}-\boldsymbol{\omega}^{(0)}_{L/I}\times(\boldsymbol{r}_c-\boldsymbol{r}_t)\bigr],
  \end{split}
  \label{eq:relative_state_def}
\end{equation}
which together form the local relative state.
Both $\boldsymbol{\rho}$ and $\dot{\boldsymbol{\rho}}$ are obtained from the MEE states of the two spacecraft through the ECI position and velocity, so the two descriptions are deterministic functions of the same pair $(\boldsymbol{x}_c,\boldsymbol{x}_t)$. For small separations, the relative motion about a circular target orbit with mean motion $n_t=\sqrt{\mu/a_t^3}$ is approximated by the linear Clohessy--Wiltshire (CW) equations \cite{clohessy1960terminal}. They serve here only as a reference for interpreting the terminal regime; training and evaluation propagate the perturbed MEE dynamics.

The two descriptions emphasize different control objectives. In the far-range regime, $\|\boldsymbol{\rho}\|$ reaches thousands of kilometers, the CW linearization is invalid, and the relative state varies rapidly within each revolution, whereas the MEE residuals describe the orbital mismatch directly. In the terminal regime, meter-level relative-position errors can correspond to very small orbital phase differences. For example, a 10-m along-track separation on the same near-circular orbit of radius $7\times10^6$~m corresponds to a phase offset of approximately $1.4\times10^{-6}$~rad, without a difference in $p$. The relative state expresses the terminal tolerances directly in meters and meters per second. These differences in scale and interpretation motivate assigning the two descriptions to separate experts in Section~\ref{sec:method}.

\subsection{Feedback Guidance Problem Formulation}
\label{sec:formulation}
At each guidance instant $t_k$, the observation $\boldsymbol{o}_k$ is constructed from the estimated states of both spacecraft and the elapsed time,
\begin{equation}
  \boldsymbol{o}_k=\boldsymbol{o}\!\left(\tilde{\boldsymbol{x}}_c(t_k),\tilde{\boldsymbol{x}}_t(t_k),t_k\right),
  \label{eq:observation}
\end{equation}
where $\boldsymbol{o}(\cdot)$ is a deterministic function of the state estimates and time, such as the global and local descriptions of Section~\ref{sec:frames}; in the nominal setting the estimates equal the true states, and estimation errors are considered only in Section~\ref{sec:res_robustness}. A guidance law is a causal mapping from the observation history to the command,
\begin{equation}
  \boldsymbol{u}_k=\pi(\boldsymbol{o}_0,\ldots,\boldsymbol{o}_k)\in\mathcal{U}.
  \label{eq:feedback_policy}
\end{equation}
Although the state pair augmented with time, $(\boldsymbol{x}_c,\boldsymbol{x}_t,t_k)$, is Markovian under the nominal dynamics, a reduced observation such as the local relative state need not be; \eqref{eq:feedback_policy} therefore admits history dependence, with memoryless and finite-window policies (Section~\ref{sec:method}) as special cases.

Let $T_{\max}=K_{\max}\Delta t$ be the mission deadline, where $K_{\max}$ is the maximum number of guidance intervals. The terminal tolerances are monitored on the integration grid: each guidance interval is divided into $M$ integration steps of length $\Delta t_{\mathrm s}=\Delta t/M$, giving the sample instants $\mathcal{T}=\{j\Delta t_{\mathrm s}:\,j=1,\ldots,MK_{\max}\}$. The rendezvous time is the first sample instant at which both terminal tolerances are met,
\begin{equation}
  \begin{split}
    t_f=\min\bigl\{t\in\mathcal{T}:\;&
      \|\boldsymbol{\rho}(t)\|_2<\varepsilon_r,\\
    &\|\dot{\boldsymbol{\rho}}(t)\|_2<\varepsilon_v\bigr\},
  \end{split}
  \label{eq:terminal_time}
\end{equation}
where $\varepsilon_r$ and $\varepsilon_v$ are the terminal position and velocity tolerances. The terminal set $\mathcal{S}$ consists of all relative states satisfying both inequalities in \eqref{eq:terminal_time}. Success is therefore declared as soon as any sample, including one inside a guidance interval, lies in $\mathcal{S}$; the mission then terminates without completing that interval. If the set in \eqref{eq:terminal_time} is empty, set $t_f=+\infty$ and declare the mission unsuccessful. For later use, the success indicator of guidance interval $k$ is $\mathbf{1}^{\mathrm{succ}}_k=1$ if $t_f\in(t_k,t_{k+1}]$ and $0$ otherwise.

For each executed guidance interval, define its end time and actual duration as
\begin{equation}
  t_{k+1}^{\mathrm{end}}=\min(t_{k+1},t_f),\qquad
  \Delta t_k=t_{k+1}^{\mathrm{end}}-t_k.
  \label{eq:executed_interval}
\end{equation}
Each interval is propagated to $t_{k+1}^{\mathrm{end}}$. If the terminal tolerances are met within an interval, that interval ends at $t_f$; otherwise, propagation continues to $t_{k+1}$ with $\Delta t_k=\Delta t$.

For a successful mission, the flight time is $t_f$ and the cumulative velocity increment is
\begin{equation}
  \Delta v=a_{\max}\int_{0}^{t_f}\|\boldsymbol{u}(t)\|_2\,\mathrm{d}t,
  \label{eq:cost}
\end{equation}
which serves as a proxy for propellant use \cite{betts2010practical}.

For the nominal problem, let $\boldsymbol{\xi}=(\boldsymbol{x}_c(0),\boldsymbol{x}_t(0))\sim\mathcal{D}$ denote the initial state pair drawn from the mission distribution specified in Section~\ref{sec:setup_missions}. The numerical values of $\varepsilon_r$, $\varepsilon_v$, and $K_{\max}$ are given in Section~\ref{sec:setup_metrics}. For a policy $\pi$, the success probability is
\begin{equation}
  P_\pi=\Pr_{\boldsymbol{\xi}\sim\mathcal{D}}\bigl\{t_f(\pi,\boldsymbol{\xi})\leq T_{\max}\bigr\}.
  \label{eq:policy_performance}
\end{equation}

\begin{problem}
  \label{prob:feedback_guidance}
  Find a guidance law $\pi$ of the form \eqref{eq:feedback_policy} that, subject to the dynamics \eqref{eq:mee_dynamics}--\eqref{eq:j2_acceleration} and the command constraint $\boldsymbol{u}_k\in\mathcal{U}$, first maximizes $P_\pi$ and then reduces the mean flight time and mean $\Delta v$ over successful missions.
\end{problem}

Success is thus the primary criterion, and flight time and $\Delta v$ are compared over successful missions. The rewards of Section~\ref{sec:training} serve as shaped surrogates for these objectives.

\section{TEMPORAL ATTENTION-ENHANCED MOE GUIDANCE ARCHITECTURE}
\label{sec:method}

\subsection{Architecture Overview}
\label{sec:overview}
\begin{figure}[!htbp]
\centering
\includegraphics[width=0.98\linewidth]{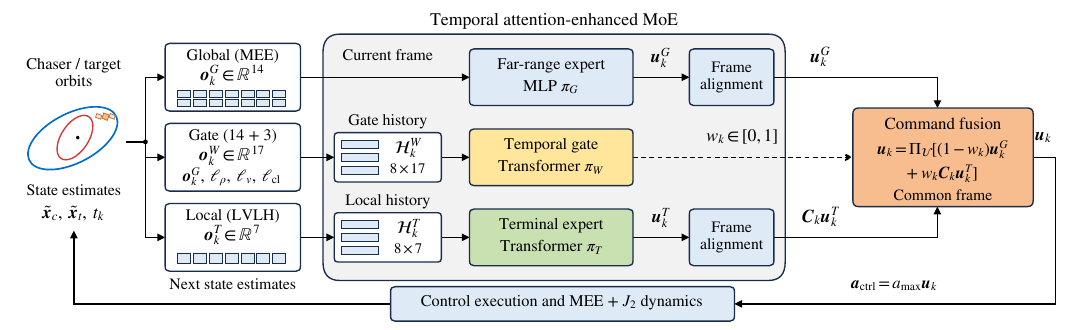}
\caption{MoE guidance architecture with temporal observations, command alignment, and thrust projection.}
\label{fig:moe_guidance_architecture}
\end{figure}

Figure~\ref{fig:moe_guidance_architecture} shows the far-range transfer expert $\pi_G$, terminal rendezvous expert $\pi_T$, and temporal gate $\pi_W$. The far-range expert uses the current global MEE observation $\boldsymbol{o}^G_k$, while the terminal expert and the gate use observation windows $\mathcal{H}^T_k$ and $\mathcal{H}^W_k$, respectively. Both windows include validity masks as defined in Section~\ref{sec:encoder} and are updated at every guidance instant, regardless of the blending weight. The deterministic guidance law is
\begin{equation}
  \begin{split}
    \boldsymbol{u}_k&=\Pi_{\mathcal{U}}\!\bigl[(1-w_k)\,\boldsymbol{u}^G_k+w_k\,\boldsymbol{C}_k\boldsymbol{u}^T_k\bigr],\\
    \boldsymbol{u}^G_k&=\pi_G(\boldsymbol{o}^G_k),\quad
    \boldsymbol{u}^T_k=\pi_T(\mathcal{H}^T_k),\\
    w_k&=\pi_W(\mathcal{H}^W_k),
  \end{split}
  \label{eq:fusion_overview}
\end{equation}
where $w_k\in[0,1]$ is the terminal expert's weight, $\boldsymbol{C}_k=[\boldsymbol{C}_{\mathrm{VNB}}^{\mathrm{RTN}}(t_k)]^{\mathrm T}$ rotates the terminal command from the chaser RTN frame into the VNB frame of $\boldsymbol{u}^G_k$, and the radial projection $\Pi_{\mathcal{U}}(\tilde{\boldsymbol{u}})=\tilde{\boldsymbol{u}}/\max(1,\|\tilde{\boldsymbol{u}}\|_2)$ enforces \eqref{eq:control_constraint} without changing the command direction. The guidance law \eqref{eq:fusion_overview} is therefore a finite-window policy of the form \eqref{eq:feedback_policy}.

\subsection{Temporal Observation Encoder}
\label{sec:encoder}
\begin{figure}[!htbp]
\centering
\includegraphics[width=0.72\linewidth]{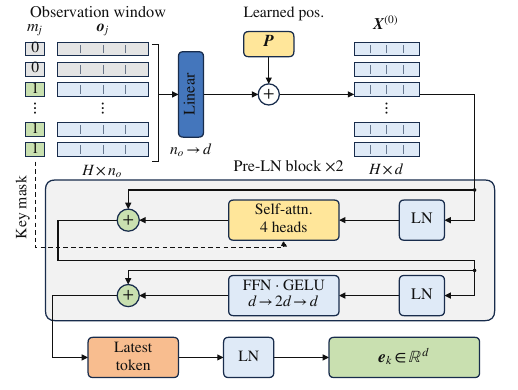}
\caption{Temporal observation encoder for the terminal expert and gate.}
\label{fig:temporal_gate_encoder}
\end{figure}

The terminal expert and the gate use Transformer encoders with the same structure and separate parameters. Define the masked observation window as $\mathcal{H}_k=((\boldsymbol{o}_{k-H+1},m_{k-H+1}),\ldots,(\boldsymbol{o}_k,m_k))$, where $\boldsymbol{o}_j\in\mathbb{R}^{n_o}$ and $m_j\in\{0,1\}$ indicates a valid frame. Invalid frames are zero-padded; at mission start only the latest frame is valid. Each observation vector is linearly embedded into $d=64$ dimensions and added to a learned positional embedding (Fig.~\ref{fig:temporal_gate_encoder}). The sequence is processed by two Transformer blocks \cite{vaswani2017attention} without dropout, with layer normalization applied before each attention and feed-forward sublayer \cite{parisotto2020stabilizing}. Each block uses four-head self-attention, in which padded frames are masked as keys, and a feed-forward network of dimensions $d\to128\to d$ with a Gaussian error linear unit (GELU) activation. No causal mask is needed because only the latest token is used and the window contains no future observations: the layer-normalized output at the latest frame gives the feature $\boldsymbol{e}_k\in\mathbb{R}^{d}$. The terminal encoder uses $n_o=7$ and $H=H_T$, while the gate encoder uses $n_o=17$ and $H=H_W$, producing features $\boldsymbol{e}^T_k$ and $\boldsymbol{e}^W_k$, respectively.

\subsection{Expert Policies}
\label{sec:experts}
The far-range expert acts on the global description of Section~\ref{sec:frames} and the terminal expert on the local one.

\subsubsection{Far-Range Transfer Expert}
\label{sec:far_expert}
The far-range expert performs orbital reconfiguration and phasing. Its observation is
\begin{equation}
  \begin{split}
    \boldsymbol{o}^G=\big[&\bar p_c,f_c,g_c,h_c,k_c,\sin L_c,\cos L_c,\\
    &\Delta\bar p,\Delta f,\Delta g,\Delta h,\Delta k,\Delta L,\tau\big]^{\mathrm T},
  \end{split}
  \label{eq:global_obs}
\end{equation}
where $\bar p=p/R_E$, the residuals are those of Section~\ref{sec:frames} with $\Delta L$ wrapped by \eqref{eq:longitude_residual}, and $\tau=(K_{\max}-k)/K_{\max}$ is the normalized remaining time. The absolute slow elements locate the chaser orbit, on which the $J_2$ drift rates and the control matrix \eqref{eq:mee_control_matrix} depend; $(\sin L_c,\cos L_c)$ encodes the orbital position without the $2\pi$ discontinuity; the residuals give the remaining shape and phase mismatch; and $\tau$ conveys the deadline $T_{\max}$. With exact observations, these features recover both MEE states, with true longitudes identified modulo $2\pi$, and the current time. The resulting Markov representation under the nominal dynamics supports a memoryless expert. The expert is a multilayer perceptron (MLP) with three 512-unit ReLU hidden layers acting on the current observation, $\boldsymbol{u}^G_k=\pi_G(\boldsymbol{o}^G_k)$.

\subsubsection{Terminal Rendezvous Expert}
\label{sec:terminal_expert}
The terminal expert drives the relative state \eqref{eq:relative_state_def} into the terminal set. With $\boldsymbol{\rho}=[x,y,z]^{\mathrm T}$, its observation is
\begin{equation}
  \boldsymbol{o}^T=\left[\frac{x}{\sigma_{xy}},\frac{y}{\sigma_{xy}},\frac{z}{\sigma_z},\frac{\dot x}{\sigma_v},\frac{\dot y}{\sigma_v},\frac{\dot z}{\sigma_v},\frac{n_{t,0}}{\sigma_n}\right]^{\mathrm T},
  \label{eq:local_obs}
\end{equation}
with $\sigma_{xy}=2000$~m, $\sigma_z=100$~m, $\sigma_v=10$~m/s, and $\sigma_n=1.2\times10^{-3}$~rad/s. These normalization scales are prescribed constants shared across missions. Here, $n_{t,0}$ denotes the target's initial mean motion. The normalized mean-motion feature $n_{t,0}/\sigma_n$ distinguishes target orbits at different initial altitudes. In the navigation-error tests it is initialized from the first measured target state. The smaller scale for $z$ increases the normalized sensitivity to cross-track errors.

The observation $\boldsymbol{o}^T_k$ is not a sufficient Markov state for the full dynamics \eqref{eq:mee_dynamics}--\eqref{eq:j2_acceleration}. Target eccentricity and orbital position affect the departure of the relative motion from the CW model, but neither is determined by the current relative state and mean motion. Thus, the same local observation can correspond to different subsequent relative motion for the same command. The terminal expert therefore faces a POMDP even when the supplied state estimates are exact.

To address this partial observability, the expert acts on the window $\mathcal{H}^T_k$ of $H_T=8$ frames, encoded by its own instance of the encoder of Section~\ref{sec:encoder}. The encoder can use recent relative-motion trends as cues to target-orbit information absent from the current observation. The window spans a duration comparable to one target orbital period for the mission class considered here (Section~\ref{sec:setup}). The feature $\boldsymbol{e}^T_k$ is mapped by three 256-unit ReLU layers to $\boldsymbol{u}^T_k=\pi_T(\mathcal{H}^T_k)$.

\subsection{Temporal Attention Gating Policy}
\label{sec:gate}

To account for the terminal expert's history-dependent command, the gate also receives an observation window. Each gate observation augments the global observation \eqref{eq:global_obs} with three local features,
\begin{equation}
  \boldsymbol{o}^W=\big[(\boldsymbol{o}^G)^{\mathrm T},\,\ell_\rho,\,\ell_v,\,\ell_{\mathrm{cl}}\big]^{\mathrm T}\in\mathbb{R}^{17}.
  \label{eq:gate_obs}
\end{equation}
The local features use numerical values of lengths in meters and velocities in meters per second:
\begin{equation}
  \begin{split}
    \ell_\rho&=\tfrac{1}{4}\big[\ln\max(\|\boldsymbol{\rho}\|,1)-8\big],\\
    \ell_v&=\tfrac{1}{4}\ln\max(\|\dot{\boldsymbol{\rho}}\|,10^{-3}),\\
    \ell_{\mathrm{cl}}&=\tfrac{1}{4}\operatorname{asinh}v_{\mathrm{cl}},\\
    v_{\mathrm{cl}}&=-\frac{\boldsymbol{\rho}^{\mathrm T}\dot{\boldsymbol{\rho}}}{\max(\|\boldsymbol{\rho}\|,\rho_{\min})}.
  \end{split}
  \label{eq:gate_local_features}
\end{equation}
Here $\boldsymbol{\rho}$ and $\dot{\boldsymbol{\rho}}$ are defined by \eqref{eq:relative_state_def}, and $\rho_{\min}=10^{-6}$~m prevents division by zero. The closing rate is positive during approach. Under exact observations, these local quantities are already determined by $\boldsymbol{o}^G$, but their scaled form makes near-target separation and relative motion explicit. The logarithms compress the range of distances and speeds, while $\operatorname{asinh}$ preserves the sign of the closing rate. The gate receives $H_W=8$ frames of \eqref{eq:gate_obs} with their validity masks.

Under exact observations, each valid gate observation determines the state pair and time, and hence the corresponding terminal observation. A gate window with $H_W\geq H_T$ therefore also determines $\mathcal{H}^T_k$, so that, with frozen experts, the gate has access to every input of both expert commands; the present design uses $H_W=H_T$.

The gate's own encoder maps $\mathcal{H}^W_k$ to $\boldsymbol{e}^W_k$, which an MLP head converts to a scalar $\eta_k$. The deterministic gate action and blending weight are
\begin{equation}
  \vartheta_k=\tanh\eta_k,\qquad w_k=\tfrac{1}{2}\left(\vartheta_k+1\right).
  \label{eq:weight_map}
\end{equation}
This scalar parameterization allows continuous blending and weights arbitrarily close to either 0 or 1. The configuration of the three networks is summarized in Table~\ref{tab:network_config}.

\begin{table}[!htbp]
  \caption{Network Configuration of the Guidance Law}
  \label{tab:network_config}
  \centering
  \renewcommand{\arraystretch}{1.15}
  \setlength{\tabcolsep}{3pt}
  \begin{tabular}{@{}llll@{}}
    \hline\hline
    Module & Input & Network & Output\\
    \hline
    $\pi_G$ & $\boldsymbol{o}^G_k\in\mathbb{R}^{14}$ & MLP $3\times512$ & $\boldsymbol{u}^G_k$\\
    $\pi_T$ & $\mathcal{H}^T_k$ ($8\times7$) & Enc.\ $+$ MLP $3\times256$ & $\boldsymbol{u}^T_k$\\
    $\pi_W$ & $\mathcal{H}^W_k$ ($8\times17$) & Enc.\ $+$ MLP $3\times256$ & $w_k\in[0,1]$\\
    \hline
    \multicolumn{4}{@{}p{0.95\columnwidth}@{}}{\footnotesize Window dimensions exclude validity bits. Enc.: two pre-LN blocks, $d=64$, four heads, FFN $64\to128\to64$ with GELU, learned positional embedding, key mask, no dropout. MLP layers use ReLU; expert outputs use componentwise $\tanh$, and the gate output uses \eqref{eq:weight_map}.}\\
    \hline\hline
  \end{tabular}
\end{table}

\section{STAGED TRAINING AND JOINT FINE-TUNING}
\label{sec:training}
The guidance law of Section~\ref{sec:method} is assembled from separately pretrained experts. With both experts frozen, the gate is trained by SAC on complete missions. The resulting guidance law then initializes end-to-end fine-tuning by proximal policy optimization (PPO) \cite{schulman2017proximal}, with the policy likelihood defined on the pre-projection fused command. The following stages specify the preparation and optimization of these modules; the effect of fine-tuning on closed-loop performance is evaluated in Section~\ref{sec:res_overall}.

All training stages and nominal evaluation use our JAX-based framework for parallel GPU simulation and training~\cite{wu2026massively}. The training settings are summarized in Table~\ref{tab:training_hyperparameters}.

\begin{table}[!htbp]
  \caption{Training Settings of the Experts, Gate, and Joint Fine-Tuning}
  \label{tab:training_hyperparameters}
  \centering
  \small
  \renewcommand{\arraystretch}{1.12}
  \setlength{\tabcolsep}{7pt}
  \begin{tabular}{@{}lcccc@{}}
    \hline\hline
    Setting & Far-range SAC & Terminal SAC & Gate SAC & Joint PPO\\
    \hline
    Parallel environments & 1280 & 480 & 480 & 960\\
    Replay/minibatch size & 8192 & 4096 & 4096 & 4096\\
    Replay capacity & $5\times10^6$ & $5\times10^6$ & $5\times10^6$ & ---\\
    Actor learning rate & $3\times10^{-4}$ & $10^{-4}$ & $10^{-4}$ & $1.5\times10^{-5}$\\
    Critic/value learning rate & $3\times10^{-4}$ & $3\times10^{-4}$ & $3\times10^{-4}$ & $10^{-4}$\\
    Discount factor & 0.995 & 0.995 & 0.999 & 0.999\\
    \hline\hline
    \multicolumn{5}{@{}p{0.97\textwidth}@{}}{\footnotesize PPO uses 384 transitions per environment per iteration, three optimization epochs, GAE parameter 0.98, clipping parameter 0.2, maximum gradient norm 0.5, KL limit 0.05, and deterministic action-change RMS limit 0.03. Its encoder learning rate is $3\times10^{-6}$; value warmup lasts 20 iterations. The joint-PPO column gives the joint fine-tuning settings; the checkpoint is selected on 600 separate validation missions.}\\
  \end{tabular}
\end{table}

\subsection{Pretrained Far-Range Expert}
\label{sec:train_far}
The far-range expert is pretrained according to the SAC training procedure and orbital-transfer curriculum described in our previous work~\cite{wu2026massively}. It enters the present framework as a pretrained policy with the observation and command interfaces specified in Section~\ref{sec:experts}. Its parameters remain fixed during handover-data collection and gate training, and are subsequently updated during joint fine-tuning.

\subsection{Terminal Expert and Handover Data}
\label{sec:train_terminal}
The terminal expert is trained on the states at which it takes over in closed loop rather than on an assumed initial distribution. The far-range expert is flown deterministically on complete transfers under \eqref{eq:mee_dynamics}--\eqref{eq:j2_acceleration}, and the first state at which each trajectory enters each of the distance bands $(0,500]$, $(500,1000]$, \ldots, $(2500,3000]$~m is recorded together with the window $\mathcal{H}^T_k$ ending at that state, including the current observation. The samples are kept in one pool per band, and a separate pool generated from independent transfers is used for evaluation. Each training episode starts from a pool sample, with the two innermost bands drawn slightly more often, and the recorded observations initialize its window $\mathcal{H}^T_0$. The expert therefore sees at handover the same history as in the complete guidance law, where $\mathcal{H}^T_k$ is updated throughout the transfer.

The expert is trained by SAC directly under \eqref{eq:mee_dynamics}--\eqref{eq:j2_acceleration}, with episodes of at most $K_{\max}^T=2000$ guidance intervals; $k$ indexes guidance intervals within that episode. With the normalized error $\delta_k=\|\boldsymbol{\Sigma}^{-1}\boldsymbol{\rho}_k\|_2+\|\dot{\boldsymbol{\rho}}_k\|_2/\sigma_v$, where $\boldsymbol{\Sigma}=\mathrm{diag}(\sigma_{xy},\sigma_{xy},\sigma_z)$ uses the scales of \eqref{eq:local_obs}, the reward of interval $k$ is
\begin{equation}
  \begin{split}
    r^T_k&=\kappa\,(\delta_k-\delta_{k+1})-\delta_{k+1}\\
    &\quad{}-c^T_v\,\Delta v_k-c^T_t+B_T\,\mathbf{1}^{\mathrm{succ}}_k.
  \end{split}
  \label{eq:terminal_reward}
\end{equation}
Here $\mathbf{1}^{\mathrm{succ}}_k$ is the success indicator of Section~\ref{sec:formulation}, equal to one if any integration sample in the interval lies in the terminal set $\mathcal{S}$, with $\varepsilon_v=2$~m/s for training. The next-state error $\delta_{k+1}$ is evaluated at $t_{k+1}$. The training thrust cost is $\Delta v_k=a_{\max}\Delta t\,\|\boldsymbol{u}_k\|_2$, with $\boldsymbol{u}_k$ denoting the executed command. The batched environment completes the full guidance interval before resetting, including the interval in which success is first detected. Evaluation instead uses the executed duration defined in \eqref{eq:executed_interval}. The coefficients are $\kappa=10$, $c^T_v=0.01$~s/m, $c^T_t=0.01$, and $B_T=5000$. The first term rewards progress toward the target, the second penalizes the remaining error, the third penalizes thrust effort, the fourth charges a fixed cost per guidance decision, and the last rewards success. Success at any integration sample marks the episode as successful; reaching the interval limit is treated as truncation, so the value is bootstrapped rather than set to zero.

\subsection{Gate Training with Frozen Experts}
\label{sec:train_gate}
With both experts frozen, the only learned quantity in the second stage is the allocation of authority. The gate is trained by SAC on complete missions drawn from $\mathcal{D}$, each ending after the first guidance interval containing a successful integration sample according to \eqref{eq:terminal_time}, or at the mission deadline $T_{\max}=K_{\max}\Delta t$. During this stage, the scalar action $\vartheta_k$ is obtained by applying $\tanh$ to a Gaussian sample with mean $\eta_k$, and the blending weight is $w_k=(\vartheta_k+1)/2$. The actor and each critic have their own copy of the encoder of Section~\ref{sec:encoder}. Each replay transition stores the complete window $\mathcal{H}^W_k$. With the weighted residual potential
\begin{equation}
  \Phi_k=\left\|\boldsymbol{D}\left[\Delta\bar p,\Delta f,\Delta g,\Delta h,\Delta k,\Delta L\right]_k^{\mathrm T}\right\|_2,
  \label{eq:mee_potential}
\end{equation}
where $\boldsymbol{D}=\mathrm{diag}(10,10,10,10,10,0.5)$, the gate reward is
\begin{equation}
  \begin{split}
    r^W_k&=c_\Phi\left(\Phi_k-\gamma_d\Phi_{k+1}\right)\\
    &\quad{}-c^W_t-c^W_v\,\Delta v_k+B_W\,\mathbf{1}^{\mathrm{succ}}_k,
  \end{split}
  \label{eq:gate_reward}
\end{equation}
with $\Phi_{k+1}=0$ at termination, $c_\Phi=5$, discount factor $\gamma_d=0.999$, $c^W_t=0.2$, $c^W_v=0.05$~s/m, and $B_W=2000$. The thrust cost $\Delta v_k$ uses the full guidance interval, as in terminal-expert training, while $c^W_t$ is charged once per guidance decision. The shaping term is potential based and therefore leaves the optimal policy unchanged \cite{ng1999policy}. It is built on the global description because, unlike the relative error, the MEE residuals describe the orbital mismatch throughout the transfer and provide a learning signal long before handover; their norm need not decrease monotonically. The encoder learning rate is one tenth of that of the policy and value heads.

\subsection{Joint Fine-Tuning}
\label{sec:train_joint}
\begin{figure}[!htbp]
\centering
\includegraphics[width=0.94\linewidth]{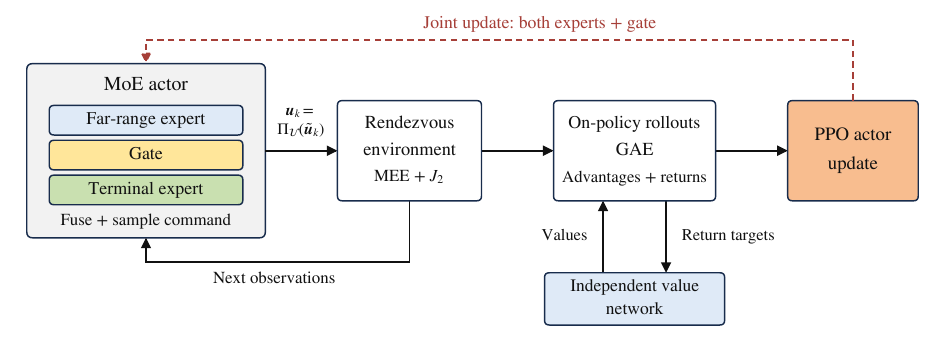}
\caption{Joint PPO fine-tuning of both experts and the gate.}
\label{fig:joint_ppo_finetuning}
\end{figure}
Figure~\ref{fig:joint_ppo_finetuning} illustrates the joint PPO update of both experts and the gate. The three networks form a joint actor with parameters $\theta$. Denote the deterministic fused command before the radial projection in \eqref{eq:fusion_overview} by
\[
  \bar{\boldsymbol{u}}_k=(1-w_k)\boldsymbol{u}^G_k+w_k\boldsymbol{C}_k\boldsymbol{u}^T_k.
\]
The expert outputs remain bounded componentwise. For exploration, the joint policy $\pi_\theta$ samples a pre-projection command $\tilde{\boldsymbol{u}}_k$ from a squashed Gaussian parameterized by $\bar{\boldsymbol{u}}_k$, with a learned exploration scale modulated by $w_k$ and a state-dependent box support enclosing the rotated expert commands. The applied command is $\boldsymbol{u}_k=\Pi_{\mathcal U}(\tilde{\boldsymbol{u}}_k)$, so \eqref{eq:control_constraint} is enforced before the dynamics and reward are evaluated. The PPO likelihood ratio compares $\pi_\theta$ with the rollout policy $\pi_{\mathrm{old}}$ using the pre-projection density; the projection is treated as part of the environment and is excluded from the likelihood. The clipped objective \cite{schulman2017proximal} therefore updates all three networks through the fused-command parameterization using returns obtained under the thrust constraint. The reward and its coefficients are the same as those of $r^W_k$ in \eqref{eq:gate_reward}. The training thrust cost uses the full guidance interval, as in expert and gate training. Evaluation stops cost accumulation at the first successful integration sample. Generalized advantage estimation \cite{schulman2016high} provides advantage estimates $\hat A_k$ and value targets $\hat R_k$. An independent value network $V_\varphi$ receives the current global and local observations and both windows through its own encoders and is fitted to $\hat R_k$, so the value loss does not update the policy.

To limit initial policy drift, the value network is first fitted for 20 PPO iterations with the actor fixed, after which both experts and the gate are updated jointly. The encoder learning rate is one fifth of that of the policy heads. Actor updates are limited by the Kullback--Leibler divergence from the rollout policy and the root-mean-square change in the pre-projection deterministic fused command. Deterministic evaluation uses $\boldsymbol{u}_k=\Pi_{\mathcal U}(\bar{\boldsymbol{u}}_k)$, as specified in Section~\ref{sec:res_baseline_configs}.

\section{NUMERICAL EXPERIMENTS}
\label{sec:results}

\subsection{Simulation Setup}
\label{sec:setup}

\subsubsection{Dynamics and Control Configuration}
\label{sec:setup_dynamics}
Both spacecraft are propagated under \eqref{eq:mee_dynamics}--\eqref{eq:j2_acceleration}. The Earth constants are $\mu=3.986004418\times10^{14}$~m$^3$/s$^2$, $R_E=6378.137$~km, and $J_2=1.082629\times10^{-3}$. The chaser thrust-acceleration magnitude is bounded by $a_{\max}=5\times10^{-4}$~m/s$^2$. Each command is held over a guidance interval of $\Delta t=800$~s. State propagation uses a fixed-step fifth-order Runge--Kutta method with Dormand--Prince coefficients, with 80 integration steps of 10~s per guidance interval. The cases with unmodeled orbital perturbations use a separate double-precision propagator with the same nominal dynamics and integration scheme.

\subsubsection{Mission Distribution and Test Sets}
\label{sec:setup_missions}
Initial conditions from the mission distribution $\mathcal{D}$ are generated by uniform draws over the intervals in Table~\ref{tab:sim_setup}. The target angular elements are obtained by adding sampled offsets to the chaser angles, whereas each eccentricity is sampled conditional on its own semimajor axis. The target semimajor-axis offset has the lower bound $\delta a_{\min}=\max\{-300~\mathrm{km},\,R_E+400~\mathrm{km}-a_c\}$. For the eccentricities, $e_{\min}=10^{-6}$ and
\[
  e_{\max}(a)=\max\!\left\{e_{\min},\,
  \min\!\left[0.1,\,1-\frac{R_E+400~\mathrm{km}}{a}\right]\right\}.
\]
These conditional bounds enforce a minimum perigee altitude of 400~km. The sampled missions cover non-coplanar elliptical transfers with arbitrary initial phase differences.

\begin{table}[!htbp]
  \caption{Initial-Orbit Sampling Ranges}
  \label{tab:sim_setup}
  \centering
  \renewcommand{\arraystretch}{1.15}
  \setlength{\tabcolsep}{3pt}
  \begin{tabular}{@{}lcc@{}}
    \hline\hline
    Element & Chaser range & Target sampling rule\\
    \hline
    $a$ (km) & $R_E+[450,\,2000]$ & $a_c+[\delta a_{\min},\,500]$\\
    $e$ & $[e_{\min},\,e_{\max}(a_c)]$ & $[e_{\min},\,e_{\max}(a_t)]$\\
    $i$ ($^\circ$) & $[40,\,80]$ & $i_c+[-0.4,\,0.4]$\\
    $\Omega$ ($^\circ$) & $[2,\,360)$ & $\Omega_c+[-1,\,1]$\\
    $\omega$ ($^\circ$) & $[0,\,360)$ & $\omega_c+[-5,\,5]$\\
    $\nu$ ($^\circ$) & $[0,\,360)$ & $\nu_c+[-180,\,180)$\\
    \hline\hline
  \end{tabular}
\end{table}

\subsubsection{Evaluation Metrics and Protocol}
\label{sec:setup_metrics}
The terminal tolerances in \eqref{eq:terminal_time} are $\varepsilon_r=10$~m and $\varepsilon_v=1$~m/s, and the mission limit is $K_{\max}=10^4$ guidance intervals (approximately 92.6~days). All full-mission comparisons use the same 600 test missions drawn from $\mathcal{D}$. The jointly fine-tuned policy (joint-FT) uses the checkpoint selected on a separate 600-mission validation set, generated with a different random seed, after approximately $6.99\times10^9$ transitions of joint fine-tuning. The training rewards and checkpoint selection used a velocity tolerance of 2~m/s; all policies are evaluated at the tighter 1~m/s without retraining or reselection. For a trajectory containing $N$ executed guidance intervals, flight time and cumulative velocity increment are computed as
\begin{equation}
  \begin{split}
    T_{\mathrm{num}}&=\sum_{k=0}^{N-1}\Delta t_k,\\
    \Delta v_{\mathrm{num}}&=a_{\max}\sum_{k=0}^{N-1}\Delta t_k\,\|\boldsymbol{u}_k\|_2.
  \end{split}
  \label{eq:numerical_costs}
\end{equation}
Here $\Delta t_k$ is defined in \eqref{eq:executed_interval}. Success rates use the full test set, and mean flight time and $\Delta v$ are computed over each method's successful missions.

\subsection{Rendezvous Performance}
\label{sec:res_baselines}

\subsubsection{Baseline Configurations}
\label{sec:res_baseline_configs}
We compare the pretrained far-range expert, five hard-switching (HS) policies, and the MoE before and after joint fine-tuning (pre-FT and joint-FT). For $d_{\mathrm{sw}}\in\{1,\,2,\,2.5,\,3,\,4\}$~km, HS updates its terminal-expert weight at each guidance instant as
\begin{equation}
  w_k^{\mathrm{HS}}=\max\!\left\{w_{k-1}^{\mathrm{HS}},\,
  \mathbf{1}_{\{\|\boldsymbol{\rho}_k\|_2\leq d_{\mathrm{sw}}\}}\right\},
  \label{eq:hard_switch_baseline}
\end{equation}
with $w_{-1}^{\mathrm{HS}}=0$. Once the threshold is reached, the terminal expert retains control for the remainder of the mission. Its observation history is updated throughout the transfer. The far-range-only and HS baselines use the same pretrained expert parameters as MoE pre-FT, whereas MoE joint-FT uses the jointly updated experts. The single-frame MLP-gate configuration of Section~\ref{sec:res_temporal_attention} uses the experts with which it was trained. All methods, including the far-range-only and HS configurations, use deterministic inference and the radial projection in \eqref{eq:fusion_overview}.

\subsubsection{Overall Rendezvous Performance}
\label{sec:res_overall}

Table~\ref{tab:rendezvous_performance} compares the guidance configurations on the same 600 test missions.

\begin{table}[!htbp]
  \caption{Rendezvous Performance on the 600 Test Missions}
  \label{tab:rendezvous_performance}
  \centering
  \small
  \renewcommand{\arraystretch}{1.15}
  \setlength{\tabcolsep}{2.5pt}
  \begin{tabular}{@{}lrrr@{}}
    \hline\hline
    Method & Success (\%) & \shortstack{Mean time\\(days)} & \shortstack{Mean $\Delta v$\\(m/s)}\\
    \hline
    Far-range expert & 0.50 (3/600) & --- & ---\\
    HS 1 km & 97.17 (583/600) & 21.948 & 716.048\\
    HS 2 km & 99.33 (596/600) & 17.262 & 607.053\\
    HS 2.5 km & 99.67 (598/600) & 15.886 & 574.828\\
    HS 3 km & 99.67 (598/600) & 15.317 & 561.481\\
    HS 4 km & 99.83 (599/600) & 13.918 & 528.241\\
    MLP gate & 99.83 (599/600) & 14.118 & 575.151\\
    MoE pre-FT & 99.83 (599/600) & 10.248 & 439.192\\
    \textbf{MoE joint-FT} & 100.00 (600/600) & \textbf{6.896} & \textbf{293.940}\\
    \hline\hline
    \multicolumn{4}{@{}p{0.98\columnwidth}@{}}{\footnotesize Counts give successful missions out of 600. All methods apply the radial thrust projection. The success tolerances are 10~m and 1~m/s. Costs stop at the first jointly successful 10-s sample and are averaged over each method's successful missions; failures are excluded from cost means. The MLP-gate row uses a single-frame gate with the attention-based terminal expert (Section~\ref{sec:res_temporal_attention}).}\\
  \end{tabular}
\end{table}

The far-range expert alone meets the terminal tolerances in only 3 of 600 missions, so its cost means are omitted. Adding the terminal expert through HS raises the success count to between 583 and 599 of 600 across the tested thresholds. The 4-km switch completes 599 missions, with a successful-mission mean flight time of 13.918~days and mean $\Delta v$ of 528.241~m/s.

MoE pre-FT completes 599 of 600 missions. On the 599 missions completed by both MoE pre-FT and HS at 4~km, the learned gate reduces mean flight time by 26.4\% and reduces mean $\Delta v$ by 16.9\%. Both configurations use the same pretrained experts, so this comparison evaluates learned blending versus the fixed distance-triggered allocation under the common thrust bound.

\begin{figure}[!htbp]
\centering
\includegraphics[width=\linewidth]{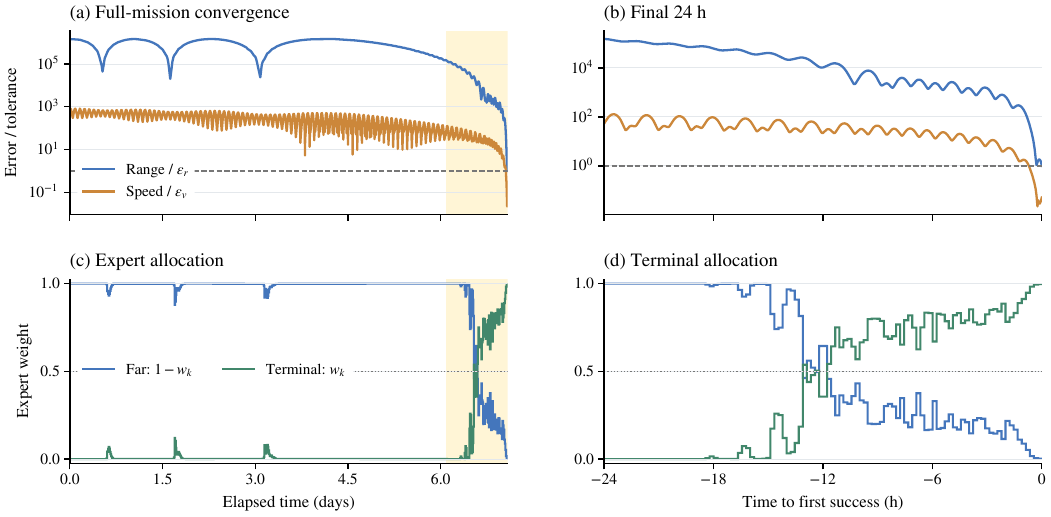}
\caption{State convergence and expert allocation in the illustrative mission: (a), (c) full mission; (b), (d) final 24~h, shaded in (a), (c).}
\label{fig:mission_coordination}
\end{figure}

Joint fine-tuning increases the success count from 599 to 600 of 600. On the 599 missions completed by both pre-FT and joint-FT, it reduces mean flight time by 32.9\% and reduces mean $\Delta v$ by 33.3\%. On the 599 missions completed by both joint-FT and HS at 4~km, the joint-FT MoE reduces mean flight time by 50.6\% and reduces mean $\Delta v$ by 44.5\%. These paired comparisons use common successful missions, while Table~\ref{tab:rendezvous_performance} reports each method's own successful-mission means.

\subsubsection{Illustrative Mission and Expert Coordination}
\label{sec:res_mission}

To illustrate the learned expert coordination, one mission from the 600-mission test set is examined in detail. It was fixed before the final evaluation rather than selected for its performance. Table~\ref{tab:illustrative_initial_orbits} lists the initial orbital elements.

\begin{table}[!htbp]
  \caption{Initial Orbital Elements of the Illustrative Mission}
  \label{tab:illustrative_initial_orbits}
  \centering
  \small
  \renewcommand{\arraystretch}{1.1}
  \setlength{\tabcolsep}{4pt}
  \begin{tabular}{@{}lrr@{}}
    \hline\hline
    Parameter & Chaser & Target\\
    \hline
    Semimajor axis, $a$ (km) & 7146.369 & 7534.813\\
    Eccentricity, $e$ & 0.003530 & 0.015251\\
    Inclination, $i$ ($^\circ$) & 68.494 & 68.765\\
    \shortstack[l]{Right ascension of the ascending\\node, $\Omega$ ($^\circ$)} & 209.693 & 209.178\\
    Argument of perigee, $\omega$ ($^\circ$) & 247.547 & 243.479\\
    True anomaly, $\nu$ ($^\circ$) & 293.598 & 140.047\\
    \hline\hline
  \end{tabular}
\end{table}

\begin{figure}[!htbp]
\centering
\includegraphics[width=0.64\linewidth]{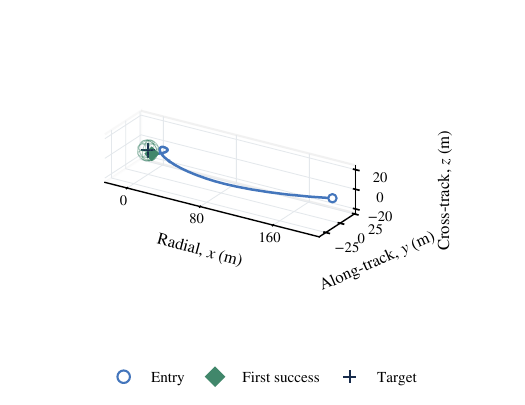}
\caption{Final 200-m approach of the illustrative mission in the target LVLH frame; the sphere marks the 10-m position tolerance.}
\label{fig:mission_geometry}
\end{figure}

Figure~\ref{fig:mission_coordination} shows the division of control authority learned by the joint-FT MoE. During the long-range transfer, the gate assigns most of the weight to the far-range expert, with brief contributions from the terminal expert. As the chaser approaches the target, the terminal contribution increases with temporary reallocations between the experts. The terminal weight remains above 0.5 after 6.593~days, when the separation is approximately 124~km. Joint fine-tuning permits this allocation beyond the pretrained terminal expert's original local domain. The allocation is learned rather than prescribed by a fixed handover radius.

With the terminal expert dominant, the final approach converges to the target in the LVLH frame (Fig.~\ref{fig:mission_geometry}). Both success tolerances are first met after 7.083~days. The accumulated $\Delta v_{\mathrm{num}}$ up to this sample is 302.561~m/s.

\begin{figure}[!htbp]
\centering
\includegraphics[width=\linewidth]{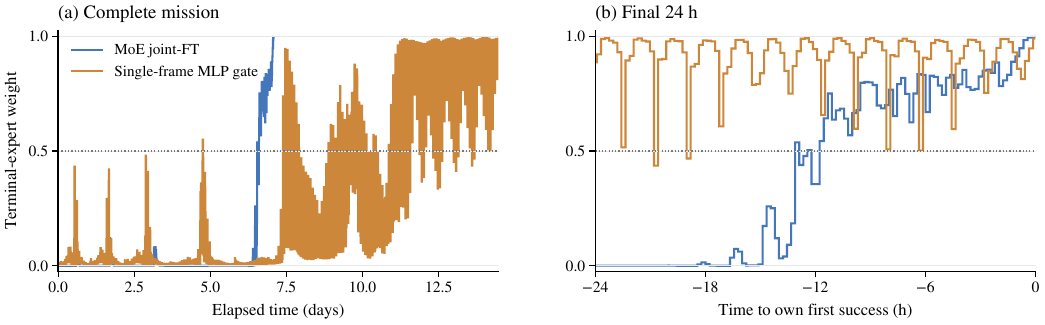}
\caption{Terminal-expert weights in the illustrative mission: (a) full mission; (b) final 24~h relative to each policy's first success.}
\label{fig:attention_gate_weights}
\end{figure}

\subsubsection{Effect of Temporal Attention}
\label{sec:res_temporal_attention}

The terminal-expert baseline is a SAC-trained MLP with three 256-unit hidden layers that takes only the current observation as input. It is evaluated on 600 terminal-phase initial conditions drawn from the held-out handover pool of Section~\ref{sec:train_terminal}; these episodes are distinct from the 600 complete rendezvous missions. With the same 10-m and 1-m/s tolerances and thrust projection, the pretrained attention-based terminal expert succeeds in all 600 cases (100.00\%), compared with 291 (48.50\%) for the MLP. The remaining 309 MLP episodes reach the 2000-interval limit. This MLP terminal policy is not used in the subsequent full-mission gate comparison.

For complete rendezvous missions, the comparison retains an attention-based terminal policy and evaluates a single-frame MLP gate alongside the temporal-gate MoE before and after joint fine-tuning. Table~\ref{tab:rendezvous_performance} summarizes the full-mission rendezvous performance of the MLP-gate, MoE pre-FT, and MoE joint-FT configurations.

The MLP-gate and pre-FT configurations each complete 599 of 600 missions, while joint-FT completes all 600. On the 599 missions completed by both MLP-gate and pre-FT, the MLP-gate configuration requires 37.8\% more flight time and 31.0\% more $\Delta v$. On the 599 missions completed by both MLP-gate and joint-FT, the corresponding increases are 105.4\% and 96.3\%, respectively.

Figure~\ref{fig:attention_gate_weights} compares the terminal weights in the illustrative mission. The MLP gate crosses 0.5 a total of 123 times, compared with 5 crossings for joint-FT. First success occurs at 14.459~days with the MLP gate and 7.083~days with joint-FT. Each trace comes from that policy's own closed-loop trajectory and ends at its first successful sample.

\subsection{Comparison of Guidance and Direct Trajectory Optimization}
\label{sec:res_conventional}
The rendezvous Q-law (RQ-law) is a Lyapunov feedback guidance law formulated in modified equinoctial elements for low-thrust rendezvous~\cite{narayanaswamy2023equinoctial}. Four solutions are compared for the initial conditions in Table~\ref{tab:illustrative_initial_orbits}: RQ-law, direct optimization initialized by RQ-law, the joint-FT MoE, and direct optimization initialized by the MoE. Each optimization retains its guidance trajectory's final time and original spacecraft states, so the two within-pair comparisons hold the transfer duration fixed. Both pairs use the nominal $J_2$ dynamics and the same $5\times10^{-4}$-m/s$^2$ thrust bound.

\begin{figure}[!htbp]
  \centering
  \includegraphics[width=\linewidth]{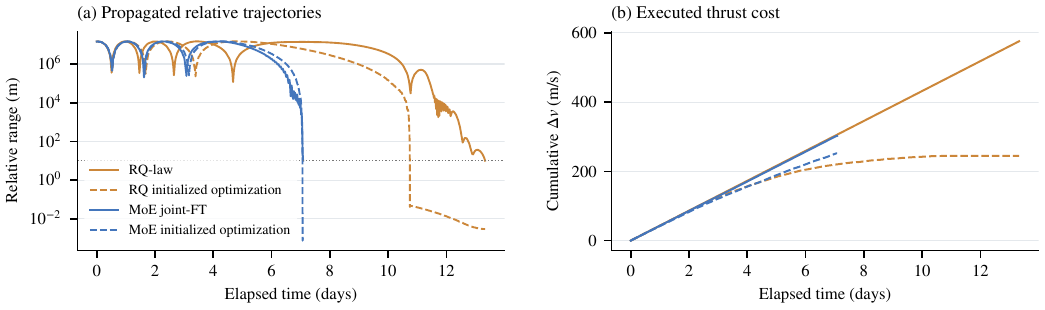}
  \caption{Guidance and optimized trajectories for the illustrative mission. Each optimized solution retains its own guidance solution's final time. Panel (a) shows relative range, and panel (b) shows cumulative executed thrust cost. Optimization costs are integrated over the entire prescribed horizon, including any samples after first entering the terminal set.}
  \label{fig:conventional_comparison}
\end{figure}

The direct optimization uses a segmented seventh-order Radau transcription in Dymos~\cite{falck2021dymos} and is solved by IPOPT~\cite{wachter2006ipopt} to minimize the time integral of thrust-acceleration magnitude. The chaser is constrained to the target state at the fixed final time. Three-axis thrust commands are held constant over 800-s intervals, with a shortened final interval. The results are summarized in Table~\ref{tab:conventional_comparison} and Fig.~\ref{fig:conventional_comparison}.

\begin{table}[!htbp]
\caption{Guidance and Fixed-Time Direct Optimization for the Illustrative Mission}
\label{tab:conventional_comparison}
\centering\small
\begin{minipage}{\columnwidth}
\setlength{\tabcolsep}{3pt}
\begin{tabular*}{\linewidth}{@{\extracolsep{\fill}}lrr@{}}
\hline\hline
Method & Time (days) & $\Delta v$ (m/s)\\
\hline
RQ-law & 13.329 & 575.830\\
RQ-law initialized optimization & 13.329 & 244.988\\
MoE joint-FT & 7.083 & 302.561\\
MoE initialized optimization & 7.083 & 252.380\\
\hline\hline
\end{tabular*}
\end{minipage}
\end{table}

The MoE reaches the joint terminal set in 7.083~days with $\Delta v=302.561$~m/s. Optimization initialized by its trajectory reduces $\Delta v$ to 252.380~m/s, a 16.6\% reduction at the same final time.

RQ-law reaches the terminal set in 13.329~days with $\Delta v=575.830$~m/s. With its trajectory as the initial guess and the final time fixed, the corresponding optimization yields $\Delta v=244.988$~m/s, a 57.5\% reduction.

\subsection{Robustness of the Learned Guidance Law}
\label{sec:res_robustness}

\subsubsection{Test Conditions and Navigation Error Model}
\label{sec:res_robust_models}
All cases use the joint-FT policy without retraining and the 600 test missions, success criterion, and mission limit of Section~\ref{sec:setup_metrics}. Flight time and $\Delta v$ are accumulated until the first successful sample or the deadline, using the thrust actually executed.

At each guidance time $t_k$, the inertial Cartesian states of spacecraft $j\in\{c,t\}$ are measured as
\begin{equation}
  \begin{aligned}
    \hat{\boldsymbol r}_{j,k}&=\boldsymbol r_{j,k}+\boldsymbol\epsilon^r_{j,k},
    &\boldsymbol\epsilon^r_{j,k}&\sim\mathcal N(\boldsymbol 0,\sigma_r^2\boldsymbol I_3),\\
    \hat{\boldsymbol v}_{j,k}&=\boldsymbol v_{j,k}+\boldsymbol\epsilon^v_{j,k},
    &\boldsymbol\epsilon^v_{j,k}&\sim\mathcal N(\boldsymbol 0,\sigma_{\mathrm{nav},v}^2\boldsymbol I_3).
  \end{aligned}
  \label{eq:robust_navigation_error}
\end{equation}
The errors are independent between spacecraft, Cartesian components, position and velocity, and guidance times. Samples are drawn at initialization and every 800~s, with the standard deviations in Table~\ref{tab:robust_environment}. Both experts, the gate, and the frame transformations use observations derived consistently from these estimates, while their histories retain the original noisy samples. True states are used for propagation and success checks; no navigation filter is modeled.

\begin{table}[!htbp]
  \caption{Navigation and Environmental Parameters}
  \label{tab:robust_environment}
  \centering
  \small
  \renewcommand{\arraystretch}{1.12}
  \setlength{\tabcolsep}{3pt}
  \begin{tabular}{@{}>{\raggedright\arraybackslash}p{0.46\columnwidth}>{\raggedright\arraybackslash}p{0.49\columnwidth}@{}}
    \hline\hline
    Parameter & Setting\\
    \hline
    Position error $\sigma_r$ & 5~m per spacecraft per axis\\
    Velocity error $\sigma_{\mathrm{nav},v}$ & 0.025~m/s per spacecraft per axis\\
    Initial epoch & 2020-01-01 00:00:00 UTC\\
    Solar-flux inputs & $F_{10.7}=\overline F_{10.7}=150$~sfu\\
    Geomagnetic input & Daily $A_p=4$\\
    Drag coefficient $C_D$ & 2.2\\
    Reflectivity coefficient $C_R$ & 1.3\\
    Area-to-mass ratio (drag and SRP) & $0.01$~m$^2$/kg\\
    \hline\hline
  \end{tabular}
\end{table}

The navigation-error reference uses the nominal $J_2$ dynamics and thrust with the measurement errors above. The unmodeled-perturbation case adds the forces described below to this reference. The eclipse and thrust-loss cases each add their respective thrust limitation to the unmodeled-perturbation case. All cases use the same noise samples for each mission and guidance time.
The navigation-error reference completes 600 of 600 missions, with mean flight time 7.155~days and mean $\Delta v=300.988$~m/s. Relative to exact-state nominal evaluation, these means are 3.8\% and 2.4\% higher. Table~\ref{tab:robustness_results} reports the success counts and successful-mission costs for every case.

\begin{table}[!htbp]
\caption{Robustness Results on the 600 Test Missions}
\label{tab:robustness_results}
\centering\small
\begin{minipage}{0.97\textwidth}
\setlength{\tabcolsep}{5pt}
\begin{tabular*}{\linewidth}{@{\extracolsep{\fill}}lrrr@{}}
\hline\hline
Case & Success (\%) & Time (days) & $\Delta v$ (m/s)\\
\hline
Nominal, exact state & 100.00 (600/600) & 6.896 & 293.940\\
Navigation reference & 100.00 (600/600) & 7.155 & 300.988\\
Unmodeled perturbations & 100.00 (600/600) & 7.157 & 301.018\\
Unmodeled perturbations + eclipse & 100.00 (600/600) & 11.194 & 355.455\\
Unmodeled perturbations + 20\% thrust loss & 100.00 (600/600) & 9.445 & 318.671\\
\hline\hline
\end{tabular*}
\par\vspace{2pt}
{\footnotesize\noindent All nonnominal cases use the same navigation errors and noise realizations. The eclipse and thrust-loss cases use the dynamics of the unmodeled-perturbation case. Costs are averaged over successful missions and use the executed thrust.\par}
\end{minipage}
\end{table}

\subsubsection{Unmodeled Orbital Perturbations}
\label{sec:res_robust_perturbations}
The unmodeled-perturbation case adds atmospheric drag, solar radiation pressure (SRP), and solar and lunar third-body gravity to the nominal $J_2$ model for both spacecraft, with the coefficients and area-to-mass ratio in Table~\ref{tab:robust_environment}. Drag uses the velocity relative to a co-rotating atmosphere, with density from NRLMSISE-00 \cite{picone2002nrlmsise} under the fixed solar and geomagnetic inputs of the table. Third-body gravity treats the Sun and Moon as point masses with positions from DE440s \cite{park2021de440}. SRP follows a cannonball model and vanishes in the Earth's shadow, defined as in Section~\ref{sec:res_robust_eclipse}; thrust remains available at the nominal $a_{\max}$ throughout this test.
The unmodeled-perturbation case completes 600 of 600 missions. The successful-mission means are 7.157~days and 301.018~m/s. On the 600 missions completed in both this case and the navigation reference, mean flight time and mean $\Delta v$ change by less than 0.1\%.

\subsubsection{Eclipse-Induced Thrust Outages}
\label{sec:res_robust_eclipse}
\begin{figure}[!htbp]
\centering
\includegraphics[width=0.70\linewidth]{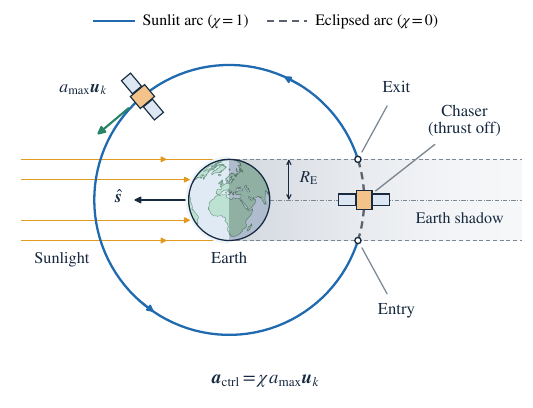}
\caption{Schematic of cylindrical Earth shadow and eclipse-induced thrust outages.}
\label{fig:eclipse_constraint}
\end{figure}

The eclipse case adds thrust outages in Earth's shadow to the unmodeled-perturbation case, retaining the nominal $a_{\max}$. Figure~\ref{fig:eclipse_constraint} illustrates the cylindrical shadow model. With $\hat{\boldsymbol s}$ the unit vector from the Earth to the Sun and $\boldsymbol r_j$ the ECI position of spacecraft $j\in\{c,t\}$, the illumination factor is $\chi_j=0$ when both
\begin{equation}
  \begin{aligned}
    \boldsymbol r_j^{\mathrm T}\hat{\boldsymbol s}&<0,\\
    \left\|\boldsymbol r_j-
      (\boldsymbol r_j^{\mathrm T}\hat{\boldsymbol s})\hat{\boldsymbol s}\right\|&<R_E,
  \end{aligned}
  \label{eq:robust_shadow}
\end{equation}
and $\chi_j=1$ otherwise. The Sun direction is evaluated from DE440s, with penumbra neglected. The applied thrust is multiplied by $\chi_c$; after eclipse exit, the currently held command resumes. The policy receives no eclipse information and retains its 800-s update interval.

\begin{figure}[!htbp]
\centering
\includegraphics[width=\linewidth]{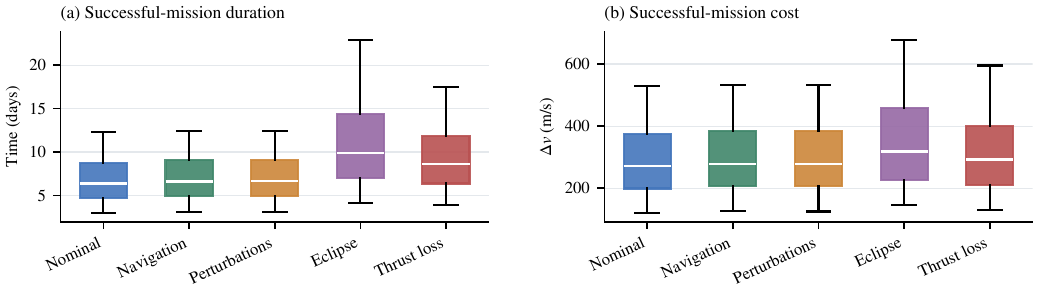}
\caption{Flight-time and executed thrust-cost distributions over successful missions for the joint-FT policy. Boxes span the interquartile range, white lines mark medians, and whiskers mark the 5th and 95th percentiles. All nonnominal cases use the same navigation-error model. The unmodeled-perturbation case is labeled Perturbations; the eclipse and thrust-loss cases add their respective thrust limitations to that case.}
\label{fig:robustness_results}
\end{figure}
The eclipse case completes 600 of 600 missions, with successful-mission mean flight time 11.194~days and mean executed $\Delta v=355.455$~m/s. The successful trajectories spend an average of 22.1\% of their duration in shadow. On 600 common successes, mean time is 56.4\% higher and mean executed $\Delta v$ is 18.1\% higher than in the unmodeled-perturbation case. $\Delta v$ is computed by integrating the magnitude of the executed thrust acceleration. During eclipse, the applied thrust is zero, so nonzero thrust commands contribute no $\Delta v$, while time in shadow remains part of the mission duration.

\subsubsection{Thrust Loss}
\label{sec:res_robust_reduced_thrust}
To simulate a 20\% thrust loss in the electric thruster, the maximum acceleration is reduced to $a_{\max,\mathrm{test}}=0.8a_{\max}=4\times10^{-4}$~m/s$^2$, with the dynamics of the unmodeled-perturbation case and continuous thrust availability. The loss affects only thrust execution; observation construction and command normalization remain unchanged. The cost expression \eqref{eq:numerical_costs} therefore uses $a_{\max,\mathrm{test}}$ in place of $a_{\max}$.
With the same navigation errors and unmodeled orbital perturbations, the thrust-loss case completes 600 of 600 missions. The mean time is 9.445~days and mean executed $\Delta v$ is 318.671~m/s. Across 600 common successes, these costs are 32.0\% higher and 5.9\% higher, respectively, than in the unmodeled-perturbation case. The increased integrated thrust cost shows that reducing the instantaneous acceleration limit does not imply a lower mission-level $\Delta v$.

All 600 missions are completed in the unmodeled-perturbation, eclipse, and thrust-loss cases. Figure~\ref{fig:robustness_results} summarizes the distributions. The tests use one noise realization per mission. Simultaneous eclipse outages and thrust loss, or larger measurement errors, are not assessed.

\section{CONCLUSION}
\label{sec:conclusion}
This paper presented a temporal attention-enhanced MoE guidance law for multi-revolution low-thrust rendezvous. A far-range expert acting on orbital elements and a terminal expert acting on the relative-state history are blended by a learned gate; the three networks are pretrained in stages and then fine-tuned end to end. On the test missions, the resulting guidance law achieves rendezvous in every case at terminal tolerances of 10~m and 1~m/s. Relative to distance-based hard switching, it approximately halves the mean flight time and lowers the mean $\Delta v$ by more than 40\%, with both the learned gate and the joint fine-tuning contributing to this improvement. For the illustrative mission, a fixed-time direct optimization leaves a moderate $\Delta v$ gap, which indicates the cost of feedback guidance relative to an offline, mission-specific solution.

Without retraining, the guidance law also completes every test mission with navigation errors alone, with additional unmodeled orbital perturbations, and with eclipse-induced thrust outages or thrust loss added separately to the unmodeled-perturbation case. These tests cover the sampled orbit distribution and the stated error and environment models; simultaneous application of the two thrust limitations, navigation-filter dynamics, and attitude and actuator errors remain for future work.

\section*{ACKNOWLEDGMENT}
The authors used AI-assisted tools to improve the language and readability of the manuscript. The authors take full responsibility for the final content.

\bibliographystyle{IEEEtaes}
\bibliography{references}

\clearpage
\section*{Author Biographies}
\begin{authorbiography}[{\includegraphics[trim={26bp 174bp 26bp 50bp},clip,width=1in,height=1.25in,keepaspectratio]{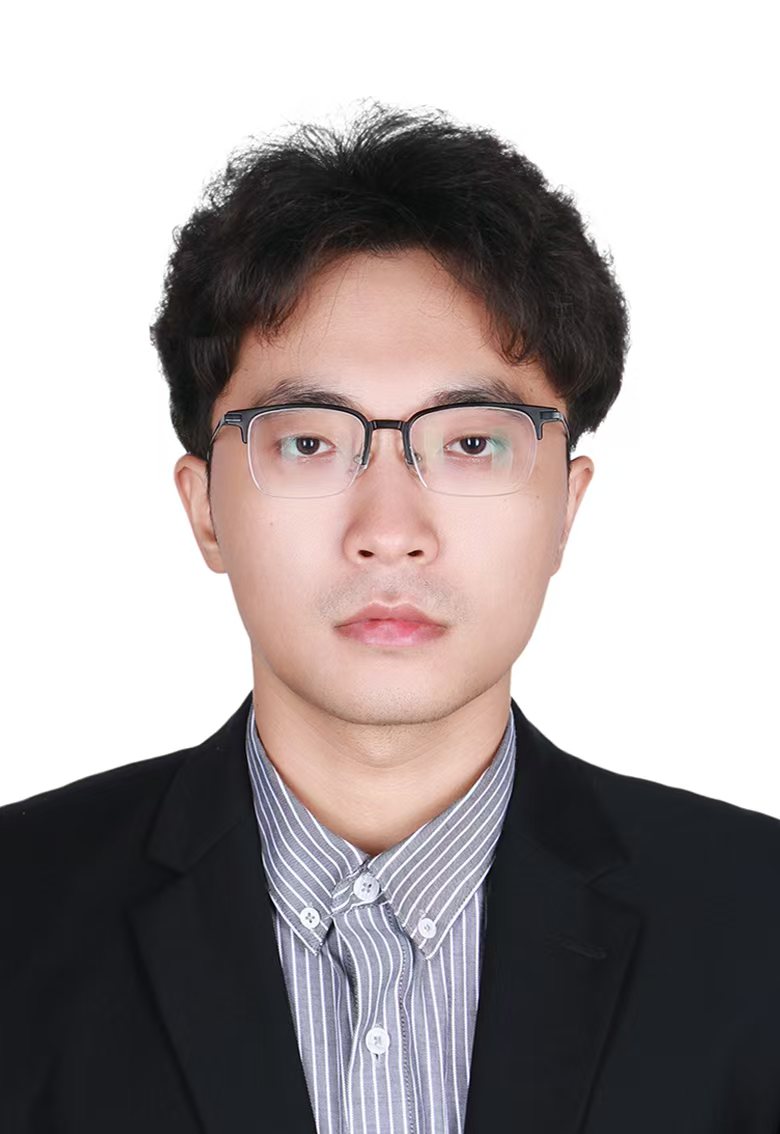}}]{Chenglong Wu}
received the B.Eng. degree in aerospace engineering from Nanjing University of Aeronautics and Astronautics, Nanjing, China, in 2024. He is currently working toward the Doctor of Engineering degree in mechanical engineering with the School of Aerospace Engineering, Tsinghua University, Beijing, China.

His research interests include reinforcement learning, high-performance computing, and low-thrust trajectory optimization.
\end{authorbiography}

\begin{authorbiography}[{\includegraphics[trim={1.5bp 3bp 1.5bp 2bp},clip,width=1in,height=1.25in,keepaspectratio]{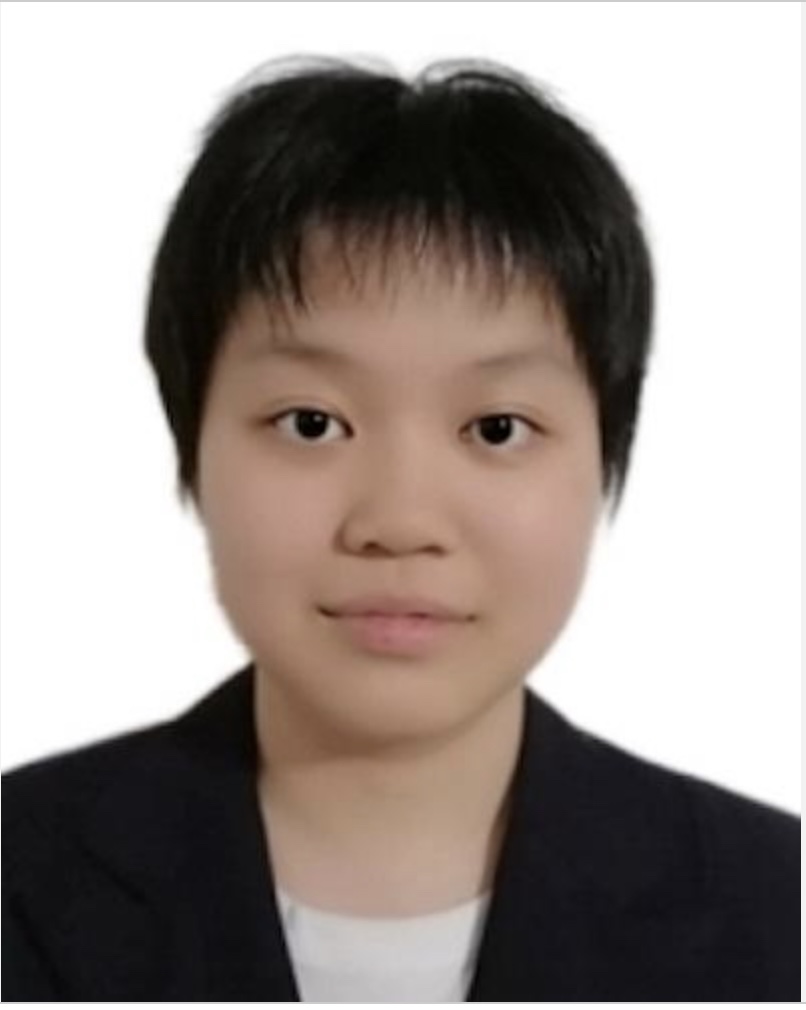}}]{Yi Zhou}
received the B.S. degree in theoretical and applied mechanics from Tsinghua University, Beijing, China, in 2025. She is currently pursuing the Ph.D. degree in aeronautical and astronautical science and technology with the School of Aerospace Engineering, Tsinghua University, Beijing, China.

Her research interests include space trajectory optimization and spacecraft guidance and control.
\end{authorbiography}

\begin{authorbiography}[{\includegraphics[trim={28bp 265bp 66bp -40bp},clip,width=1in,height=1.25in,keepaspectratio]{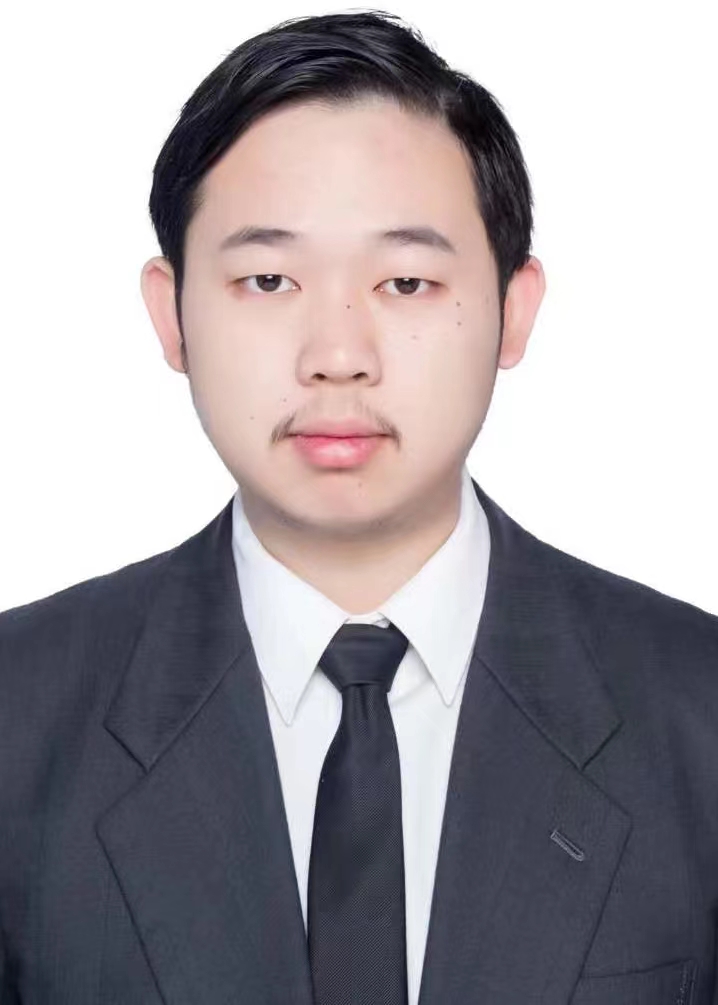}}]{Hongxu Zhu}
received the B.S. degree from the School of Astronautics, Beihang University, Beijing 102206, China, in 2017, the M.S. degree from the College of Automation Engineering, Nanjing University of Aeronautics and Astronautics, Nanjing 211106, Jiangsu, China, in 2020, and the Ph.D. degree from the School of Aeronautics and Astronautics, Shanghai Jiao Tong University, Shanghai 200240, China, in 2025.

He is currently a postdoctoral researcher with the School of Aerospace Engineering, Tsinghua University, Beijing 100084, China. His research interests include differential games, multi-agent reinforcement learning, and intelligent guidance and control.
\end{authorbiography}
\begin{authorbiography}[{\includegraphics[trim={7.086614bp 22.677165bp 7.086614bp 8.503937bp},clip,width=1in,height=1.25in,keepaspectratio]{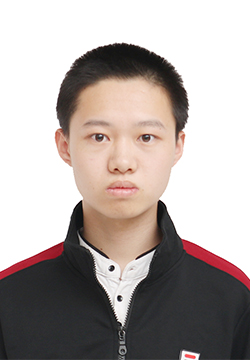}}]{Jialong Song}
received his bachelor's degree in aerospace engineering from Tsinghua University, Beijing, China, in 2023. He is currently a Ph.D. student with the School of Aerospace Engineering, Tsinghua University, Beijing, China.

His research interests include trajectory optimization.
\end{authorbiography}

\begin{authorbiography}[{\includegraphics[trim={37.9bp 0bp 37.9bp 0bp},clip,width=1in,height=1.25in,keepaspectratio]{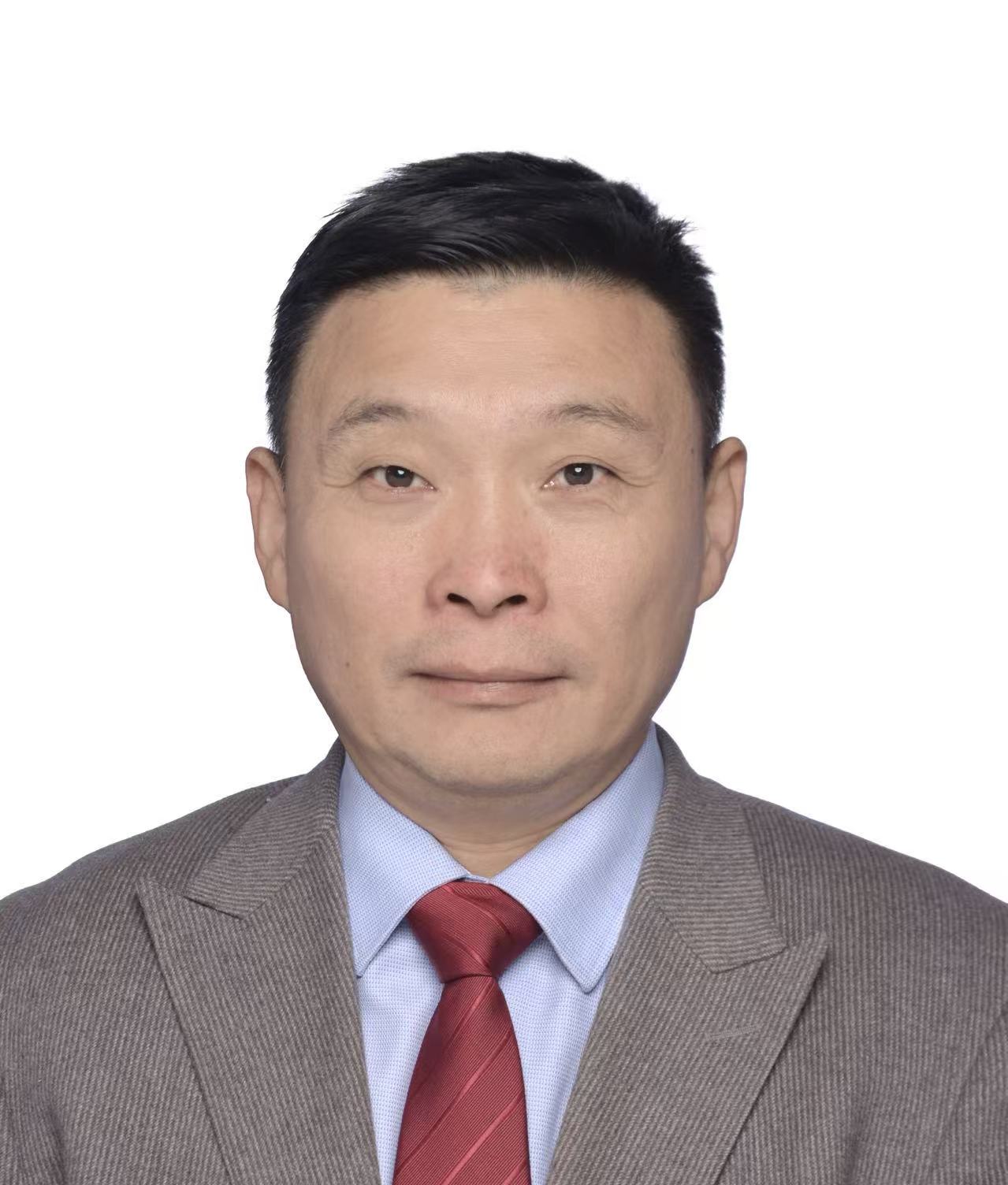}}]{Hexi Baoyin}
received the Ph.D. degree in space system engineering from the Harbin Institute of Technology, Harbin, China, in 1999.

He is currently a Professor with the School of Aerospace Engineering, Tsinghua University, Beijing, China. His research interests include space system engineering, interplanetary mission analysis, nonlinear dynamics, stability theory, and asteroid dynamics and control.
\end{authorbiography}
\end{document}